\documentclass[journal,twoside,web]{color}
\usepackage{generic}
\usepackage{cite}
\usepackage{amsmath,amssymb,amsfonts}
\usepackage{bbold}
\usepackage{algorithmic}
\usepackage{graphicx}
\usepackage{textcomp}
\usepackage{steinmetz}
\usepackage{comment}
\def\x{{\mathbf x}}
\def\L{{\cal L}}
\newcommand{\R}{\mathbf{R}}

\newcommand{\F}{\mathcal{F}}
\newcommand{\vc}[1]{\mathbf{#1}}
\newcommand{\defsign}{\triangleq}
\renewcommand{\Re}{\operatorname{Re}}
\renewcommand{\Im}{\operatorname{Im}}
\newcommand{\diag}{\operatorname{diag}}
\newcommand{\vr}{\vc{r}}
\newcommand{\vk}{\vc{k}}
\newcommand{\vv}{\vc{v}}
\newcommand{\delv}{\boldsymbol{\Delta}\vv}
\newcommand{\rect}{\operatorname{rect}}
\newenvironment{densitemize}
{\begin{list}{$\bullet$ \hfill}{
			\setlength{\leftmargin}{\parindent}
			\setlength{\parsep}{0.07\baselineskip}
			\setlength{\itemsep}{0.7\parsep}
			\setlength{\labelwidth}{\leftmargin}
			\setlength{\labelsep}{0em}}}{\end{list}}
\def\BibTeX{{\rm B\kern-.05em{\sc i\kern-.025em b}\kern-.08em
    T\kern-.1667em\lower.7ex\hbox{E}\kern-.125emX}}
\begin{document}
\title{Circumventing the resolution-time tradeoff\\ in Ultrasound Localization Microscopy\\ by Velocity Filtering}
\author{Ufuk Soylu and Yoram Bresler
\thanks{This work was supported in part by the National Science Foundation
(NSF) under Grant IIS 14-47879, and by  US Army MURI Award W911NF-15-1-0479.}
\thanks{Ufuk Soylu and Yoram Bresler are with the Department of Electrical and Computer Engineering, and the Coordinated Science Laboratory, University of Illinois at Urbana-Champaign, Urbana, IL 61801 USA. 
(e-mail: usoylu2@illinois.edu; ybresler@illinois.edu).}
}

\maketitle

\begin{abstract}
Ultrasound Localization Microscopy (ULM) offers a cost-effective modality for microvascular imaging by using intravascular contrast agents (microbubbles). However, ULM has a fundamental trade-off between acquisition time and spatial resolution, which makes clinical translation challenging. In this paper, in order to circumvent the trade-off, we 
introduce a spatio-temporal filtering operation dubbed velocity filtering, which is capable of separating contrast agents into different groups based on their vector velocities thus reducing interference in the localization step, while simultaneously offering blood velocity mapping at super resolution, without tracking individual microbubbles.  As  side  benefit,  the  velocity filter provides  noise  suppression  before  microbubble  localization that could enable substantially increased penetration depth in tissue typically by 4cm or more.
We provide a theoretical analysis of the performance of velocity filter. Numerical experiments confirm that the proposed velocity filter is able to separate the microbubbles with respect to the speed and direction of their motion. In combination with subsequent localization of microbubble centers, e.g. by matched filtering, the velocity filter  improves the quality of the reconstructed vasculature significantly and provides blood flow information. Overall, the proposed imaging pipeline in this paper enables the use of higher concentrations of microbubbles while preserving spatial resolution, thus helping circumvent the trade-off between acquisition time and spatial resolution. Conveniently, because the velocity filtering operation can be implemented by fast Fourier transforms (FFTs) it admits fast, and potentially real-time realization. We believe that the proposed velocity filtering method has the potential to pave the way to clinical translation of ULM.
\end{abstract}

\section{Introduction}
\label{sec:intro}
\subsection{Ultrasound Localization Microscopy (ULM) and its Limitations}
Diagnostic Ultrasound Imaging (DUI) has four main advantages over other biomedical imaging methods that make it indispensable:  safety; cost-effectiveness; excellent temporal resolution; and portability. As a consequence, DUI is a first choice for a variety of clinical applications \cite{book}. One such application is high-fidelity microvascular imaging. This is of great interest for two reasons. First, it provides diagnostics for diseases that modify the microvascular blood circulation. This is  crucial in oncology, where the tumor's vascular structure
contains critical diagnostic information  \cite{padhani2005angiogenesis} and for monitoring therapy response \cite{jia2016three}
. Second, it provides fundamental understanding of processes that modify the microvascular blood circulation such as brain activity \cite{deffieux2018functional}.  

DUI offers a cost-effective modality for microvascular imaging by using intravascular contrast agents. However, its performance suffers from a fundamental trade-off between spatial resolution and penetration depth: high resolution requires the use of higher excitation frequency, which comes at the cost of penetration depth, because higher frequencies attenuate more.

In  Ultrasound Localization Microscopy (ULM) \cite{errico2015ultrafast,viessmann2013acoustic,christensen2014vivo}, the trade-off between spatial resolution and penetration depth is avoided by using  ideas from Photoactivation Localization Microscopy (PALM) and Stochastic Optical Reconstruction Microscopy (STORM) \cite{betzig2006imaging,rust2006sub}. The principle of ULM is to localize the centers of individual microbubbles injected into the blood stream in a sequence of snapshot-like ultrasound images (time frames), and superimpose them to construct the aggregate super-resolved image. 

Each microbubble 
creates in each time frame a response equal to the point spread function (PSF) of the imaging system. As long as the response to one bubble is well-separated from that of other bubbles, the accuracy with which the center of the bubble response can be localized is only limited by the signal to noise ratio (SNR), rather than by the width of the PSF. It follows that well-separated bubbles can be super-resolved. Now, considering the motion of the bubbles between time frames due to blood flow, it follows that when multiple time frames with such localized bubble centers are superimposed, the resulting image will contain precisely localized bubble centers at distances well below the classical Rayleigh criterion; hence, the image showing the vasculature with the flowing bubbles will be super-resolved.
Additional information, about the flow velocity, can be extracted by tracking the motion of individual bubbles between time frames.

Hence a standard processing pipeline of  ULM \cite{song2017improved} after microbubble signal extraction starts with localization of microbubble centers followed by a tracking step. The last step is to accumulate the recovered bubble centers over a time interval, producing the final image.

While ULM circumvents the fundamental tradeoff between resolution and penetration depth of conventional DUI, and it can increase the resolution by up to tenfold \cite{errico2015ultrafast}, it suffers from another tradeoff that limits its performance: a tradeoff between spatial resolution and acquisition time. A long acquisition is required to cover the vessel structure by a sufficiently dense collection of detected bubble centers. The acquisition can be shortened if one can increase the bubble concentration. However, doing so leads to overlapping bubble responses in individual time frames, whose separation and localization accuracy is once again subject to the Rayleigh criterion, thus greatly reducing or even eliminating the super-resolution effect. This trade-off between acquisition time and spatial resolution in ULM hinders its clinical translation. Our goal in this paper is to circumvent this tradeoff, and to provide a performance analysis and prediction for the proposed method.

\subsection{Related Work}
 A significant research effort has been devoted to develop  methods more resistant to the trade-off in ULM. Bar-Zion et al. \cite{bar2018sushi}  rely on sparse recovery techniques.  Solomon et al. \cite{solomon2019exploiting} and Tang et al. \cite{tang2020kalman}  use Kalman filtering to track bubble motion, and thus improve their localization accuracy. Sloun et al. \cite{van2018super} use deep learning to help separate significantly overlapped bubble responses. However, these methods only provide limited improvement.

Huang et al. \cite{huang2020short} introduce a fundamentally new idea of separating microbubble responses into groups by an additional signal processing step. They use a cone-shaped filter in the spatiotemporal frequency domain to separate spatially overlapping microbubble signals into different sub-populations. Subsequently, localization and tracking are performed for each sub-population separately,  and then the recovered bubble centres are accumulated  over all sub populations over the time frames in the acquisition interval. 

The method proposed in this paper builds on the filtering idea proposed by Huang et al. \cite{huang2020short}. Their imaging pipeline is adopted, but with the tracking step removed. On the other hand, the proposed method differs from \cite{huang2020short} in several important ways: (i) the type of the filter employed, fully accounting, for the  first time, for the spatio-temporal structure of the ULM data; (ii) unlike \cite{huang2020short}, our filtering operation does not result in any microbubble signal loss; (iii) our filter  fully separates microbubble signals based on both speed and direction of motion, improving resolution of dense or crossing vessels; (iv) the method provides directly the microbubble velocity information eliminating the need for the separate tracking step in \cite{huang2020short}; (v) we provide a complete analysis of our proposed filter's effect on microbubble responses, and their localization by matched filtering.

Lastly, the problem studied  here is related to classical radar and sonar moving target indication (MTI) or localization. MTI methods use time delays and Doppler shifts in an optimization formulation to solve for an estimate \cite{yang2017moving}. Importantly, they address a single target or just a few. In contrast, our method doesn't use time delays or Doppler shifts, instead  relying on the 3D spatiotemporal characteristics of ULM data, and is not restricted by the number of moving targets (microbubbles).


\subsection{Contributions of the Paper}
In this paper, we tackle the trade-off in ULM. Our main contribution is a filtering method that  separates the microbubbles with respect to their vector velocity, i.e., the speed and direction of  motion. The standard processing pipeline is modified by introducing the filtering step before localization. The filter attenuates and distort the responses to microbubbles with velocities different to the selected velocity, while leaving intact the responses to microbubbles with the selected velocity. Because of the typical velocity variation between microbubbles (even those flowing in a single vessel have a typical parabolic speed distribution), the velocity-selective filtering results in effectively reduced microbubble density at each selected velocity. The trade-off between spatial resolution and acquisition time in ULM is thus circumvented by using the filter before localization.
  
The  approach is a simple way to exploit microbubble dynamics and incorporate temporal information into localization. By combining velocity-selective filtering with spatial localization, this method provides microbubble separation not just in the two spatial dimensions, but in the four-dimensional (4D) velocity-position phase space. The velocity information for each localized bubble also provides velocity mapping at super resolution, unlike previous methods that require tracking individual detected bubbles. When used with a one-dimensional (1D) transducer array that typically has little resolution in the transverse plane, the proposed approach enables immediate separation of bubbles in overlapping vessels, as long as their flow velocities differ. As a side benefit, the pre-detection narrow-band spatio temporal filtering by the velocity filter provides noise suppression before microbubble localization that could enable substantially increased penetration depth in tissue by  4 cm or more.

Thus, the paper's contributions can be outlined as follows.
\begin{itemize}
    \item A filtering operation able to separate the microbubbles with respect to their vector velocities.    
    \item Localizing microbubbles simultaneously in 4D velocity-position phase space.
    \item Simultaneously capturing flow information and a super-resolved vessel map without any tracking or an extra step.
    \item Potential for significantly increased penetration depth in tissue, e.g, by about 4cm for a typical case.
    \item A quantitative theoretical analysis providing understanding, prediction of the effect of design choices and parameters, and degree of improvement over standard ULM.
    \item Numerical experiments confirming the theory and evaluating the proposed approach in a realistic ULM system.
\end{itemize}

The rest of the paper is organized as follows: 
Section \ref{sec:problem} introduces 
a mathematical model for ULM, and states the velocity filtering problem. Section \ref{proposedmethodchapter} provides a theoretical analysis of the proposed filtering operation enabling a quantitative understanding of the operation and its impact on the images produced. Section \ref{numericalexpchapter} presents  numerical experiments to verify  the theoretical analyses  with realistic PSFs. Moreover, the proposed filter is tested on realistic simulations to demonstrate and quantify its potential usefulness in practice. Last, Section \ref{chapter6} offers conclusions.

\section{Background and Problem Statement}
\label{sec:problem}
\subsection{Notation}
In the paper, $u$ is a scalar, $\vc{u}$ is a vector, $i$ is the imaginary unit,  and $f(\vc{r})$ is a scalar function of the spatial coordinates vector $\vc{r} = [x,z]^T$, where $x$ and $z$ are the lateral and depth ultrasound image coordinates. Function $f(t)$ is a scalar function of time index t, $F(\vc{k},t)$ and $F(\vc{r},\Omega)$  are the Fourier transforms of $f(\vc{r},t)$ with respect to $\vc{r}$ and $t$, respectively, with
$\vc{k}$ and $\Omega$ the corresponding radian spatial and temporal frequencies. Similarly, $F(\vc{k},\Omega)$ is the Fourier transform of $f(\vc{r},t)$ with respect to both $\vc{r}, t$.  The dot product between vectors $\vc{a}$ and $\vc{b}$ is denoted by $\vc{a}\cdot\vc{b}$, and $\delta$ is the Dirac delta. 

\subsection{Measurement Setup}
\label{measurementsetupforsim}
We introduce a typical ultrasound imaging setup  used in the analysis and in the numerical experiments. See, e.g., \cite{book} for a thorough exposition of the principles of ultrasound imaging.
\subsubsection{Ultrasound Array and Initial Image Frame Formation}
\label{UAandIIFF}
Considering the most common ultrasound array in DUI, a 1D linear ultrasound array of piezoelectric transducers (array elements) is used to create a brightnes mode (B-Mode) 2D image. Since a linear array offers little resolution in the elevation direction, all scatters along the elevation direction in the region illuminated by the array will be projected to the image plane. The array is operated in pulse mode, using an RF signal modulated by short pulses to excite the array elements. The acoustic wave is steered on transmit by applying different delays at the array elements. 
For each transmitted pulse, delay and sum (DAS) receive beamforming is used to form an image.
%
As is common practice in ULM \cite{errico2015ultrafast,song2017improved}, which requires the acquisition of multiple image frames at high speed, we use in our model and numerical experiments 
coherent plane-wave compounding
\cite{montaldo2009coherent}.
In this method, the transmit beam for a few consecutive transmit pulses is steered to create plane waves at different angles, and the beamformed images produced 
after each pulse are
combined coherently to produce high-quality B-mode image frames, with typical frame rates of hundreds per second.   

We model the ultrasound imaging system as a linear shift-invariant system characterized by its point spread function (PSF) $g(\vc{r})$. Because DUI operates in the near-field, the actual response to a point target is shift-varying.  However, shift-invariance holds approximately in a local sense. This justifies our analysis when microbubbles move a short distance in the selected time windowing duration.
For example, for a time window of 1 second and typical microbubble maximum speed in the microvasculature of 10 mm/sec \cite{ivanov1981blood}, the relevant local neighborhood has a radius of $5mm$, for which the shift-invariance approximation is reasonable.

The PSF in B-Mode US imaging depends on the display mode selected:
 i) Pre-envelope mode is the direct result of receive beamforming (with coherent plane wave compounding).
  ii) Post-envelope imaging mode removes the highly oscillatory term facilitating human interpretation.

\textit{Post-Envelope PSF:} After envelope detection, the PSF of a linear ultrasound system is usually approximated as a Gaussian function. In our analysis we simplify the generally non-isotropic response to the circularly symmetric
\begin{align}
\label{basebandPSF}
    g_e(\vc{r})  =  1/(2 \pi  \sigma_r^2)
\exp\left(-\|\vc{r} \|^2/(2\sigma_r^2) \right),
\end{align}
where $\sigma_r$ is the characteristic width of the $PSF$. In the numerical experiments however, the actual, non-Gaussian near-field PSF of the system is present.

\textit{Pre-Envelope PSF:} Before envelope detection, there is an additional oscillatory factor in the axial coordinate at the  excitation wave's center frequency: 
\begin{align}
\label{passbandPSF}
g(\vc{r}) &= g_e(\vc{r}) \cos\left(  2 \pi z/ \lambda \right)
\end{align}
\subsubsection{Localization}
\label{basiclocamethod}
In this paper, we  use a simple method for the ULM localization step - the classical  \textit{``Matched Filter,''} which is optimal for the localization of an isolated pulse on a white Gaussian noise background. 
This is implemented by forming a cross-correlation map between the ULM data and the (local) PSF, and thresholding and local peak detection. 
Our velocity-selective filter will attenuate and distort undesired microbubble signals, thus reducing the peak value of their correlation with the PSF, and leading to their rejection by the thresholding step. This will reduce the effective microbubble density in the localization step and reduce or eliminate overlapped responses, improving localization accuracy.

%

Any other localization method such as one based on sparse recovery or neural network  can be swapped into the same imaging pipeline.  Because the performance of all localization methods degrades with overlapping signals, they would all benefit significantly from 
pre-processing by the velocity filter.
\subsubsection{3D Blood Flow Model}
\label{bloodflowmodel}
Assuming fully developed laminar flow, the speed $v(\rho)$ of fluid flow in a vessel (pipe) with a circular cross-section is  
the parabolic speed profile \cite{hoskins2000haemodynamics}
\begin{align}
\label{parabolicformualspeed}
  v(\rho) = \left(1 - \rho^2/R^2\right) v_{0}  
\end{align}
where $v_{0}$ is the speed on the center line of vessel and $\rho$ is the distance from it,
and $R$ is the radius of the vessel. The microbubbles are assumed to be distributed with a uniform concentration $C_{MB}$ (bubbles/unit volume) throughout the vessel,  each moving at the velocity of the bloodflow at its location.

However, recall that in the 2D images formed using a 1D array all bubbles inside a 3D vessel will be projected onto the 2D image plane. On this plane, their density will be far from uniform, and their speeds will no longer be governed by the parabolic speed profile. This will affect the results of any ULM method, and will be taken into account in our analysis.

\subsection{The Trade-off: Spatial Resolution vs Acquisition Time}

The inherent trade-off in ULM depends on four essential parameters: microbubble concentration
; flow rate
 (blood volume 
per unit time); acquisition time 
i.e., the total time required by the imaging operation; and spatial resolution, for which we offer, in the ULM context, the following two definitions.
\begin{itemize}
    \item [Res1] \textit{Vessel Boundary Localization Accuracy:} the deviation between  estimated 
    and true blood vessel boundaries.
    \item [Res2] \textit{Smallest Identifiable Vessel Diameter:}  the smallest vessel diameter that can be reconstructed as a complete vessel distinct from the background.
\end{itemize}

Res1 is directly related to the localization accuracy of individual microbubbles close to the vessel boundaries. For an isolated microbuble, and for a given imaging setup -- array imaging parameters and signal-to-noise ratio (SNR) -- Desailly et al. \cite{desailly2015resolution} derive lower bounds on the standard deviation of the microbubble localization error, which can be used as theoretical limits for Res1.
However, these bounds can only be achieved when the microbubbles are well separated, such that the cross correlation of their responses in the beamformed image with the PSF do not overlap. In the case of overlap, these theoretical bounds completely breakdown, as the error becomes dominated by bias rather than variance. As the bubble  separation is reduced further, in the presence of realistic noise or modeling errors, they longer be separated in the beamformed image. This is a fundamental limit that applies to any localization method (including, e.g., sparsity or neural network-based), as the sample complexity or required SNR goes from polynomial to exponential in the number of overlapping PSFs at the diffraction limit \cite{chen2020algorithmic}.

Focusing instead on the resolution metric Res2 in ULM, Hingot \emph{et al} \cite{hingot2019microvascular}  derive a lower bound on the acquisition time $Tacq$ to reconstruct a vessel of diameter $d$ in terms of the microbubble uniform concentration $C_{MB}$,  and flow rate $Q$,
\begin{align}
\label{timeacfor}
    T_{acq}\geq [(Q/d) C_{MB} I_{pix}]^{-1}
\end{align}
where  $I_{pix} \ll d$ is ultrasound image pixel size (side length), typically chosen to be similar to the resolution limit Res1. 
To interpret \eqref{timeacfor}, it is combined with 
the assumption of blood flow with a parabolic velocity profile and, following Poiseuille's law, flow rate $Q \propto d^4$.
 The trade-off between spatial resolution and acquisition time then becomes apparent. 
 
 Improving spatial resolution Res2, i.e. decreasing $d$, leads to longer $T_{acq}$ since \eqref{timeacfor} has $Q/d \propto d^3$ factor in its denominator. 
 To decrease $T_{acq}$ one could  increase the microbubble concentration $C_{MB}$.
 However, 
 this would result in overlapping microbubble responses, significantly degrading localization accuracy and hence, degrading spatial resolution metric Res1. 
\subsection{Velocity Filter in ULM}
Clearly, to circumvent the trade-off between spatial resolution and acquisition time in ULM one must overcome the problem of overlapping bubble responses in the beamformed image. As described in Sec.~\ref{sec:intro}, 
we introduce a filtering operation, which we call \textit{``Velocity Filtering,''} to separate the microbubbles by both speed and direction into different groups, each of which will have lower density and therefore reduced response overlap, thus allowing the localization accuracy to approach its theoretical, single bubble limit. 
%

 Denoting the time sequence of beamformed images by $b(\vc{r},t)$, and its processed version 
 by $\phi(\vc{r},t; \vc{v_f})$,
 the filtering operation by the velocity filter tuned to select a velocity $\vc{v_f}$ is
\begin{align}
    \phi(\vc{r},t; \vc{v_f}) &=  b(\vc{r},t) \ast_{3D} h(\vc{r},t; \vc{v_f}) \nonumber 
\end{align}
where the convolution $\ast_{3D}$ is over variables $\vc{r}$ and $t$. Then, considering an image sequence with a single microbubble moving at velocity $\vc{v}$, the problem of velocity filtering is to design a filter $h(\vc{r},t; \vc{v}_f)$ with the following three desired properties:

\begin{itemize}
    \item[(P1)] When $\vc{v} = \vc{v_f}$, $\phi(\vc{r},t,\vc{v_f})$ is equal to $ b(\vc{r},t)$. 
    
    \item[(P2)] When $\vc{v} \not = \vc{v_f}$:
    \begin{itemize}
    \item[(a)] $\phi(\vc{r},t; \vc{v_f})$ is equal to an attenuated and distorted version of $ b(\vc{r},t)$; and
    
    \item[(b)] the motion in $\phi(\vc{r},t; \vc{v_f})$ is the same as in $b(\vc{r},t)$.
    \end{itemize}
    \item[(P3)]  Attenuation and distortion increase with $\| \vc{v}- \vc{v_f} \|$.  
\end{itemize}

Why the three properties are desirable, becomes clear by considering their effect on the localization method described in Sec. \ref{basiclocamethod}. 
In particular, (P1) would allow the basic localization method to successfully localize microbubbles moving at the selected velocity $\vc{v_f}$.  On the other hand, the attenuation and distortion of microbubbles moving at a different velocity per (P2-a) will enable their elimination by adjusting the threshold in the matched filter detector. By (P2-b), even those bubbles that are not sufficiently attenuated to be eliminated, will not result in incorrect vessel boundary estimates. Moreover, by (P3), the larger the deviation of the microbubble velocity $\vc{v}$ from $\vc{v_f}$, the greater the suppression of this microbbuble. 

\subsection{Assumptions}
\label{assumptionssec}
For our theoretical analysis of the effects of the velocity filter we make the following assumptions, of which the fourth has been discussed earlier:    
\begin{densitemize}
    \item Microbubbles are modeled as point scatters since their sizes $\sim 1-5 \mu m$ in radius are at least an order of magnitude smaller than both  pixel size and the width of the PSF in any biomedical ultrasound imaging system.
    
    \item
    There is one microbubble in the scene. Because velocity filtering is linear, this is not restrictive. 
    
    \item The microbubble,  is moving at a constant velocity $\vc{v} = (v_x, v_z)$ projected onto the $(x,z)$ plane. 
    Thanks to the time windowing operator that we introduce, for sufficiently short window duration, this is a good approximation.
          Then, the microbubble position projected  onto the $(x,z)$ plane is $\vc{s}(t) =  \vc{r_0}+\vc{v}t$, where $\vc{r_0}$ 
          is its initial position. 
       
    \item The imaging system is assumed to be (locally) shift-invariant with impulse response $g(\vc{r})$.   
\end{densitemize}
It follows that the beamformed image of the moving microbubble, to which we refer as the spatiotemporal ULM data, is
\begin{align}
\label{bubblesignal}
    b(\vc{r},t) = g(\vc{r} - (\vc{r_0}+\vc{v}t))
\end{align}
where  $(\vc{r},t)$ represents lateral position $x$, depth $z$,  and time $t$.

\section{Proposed Method}
\label{proposedmethodchapter}
\subsection{Approach}
\label{approachsec}
\begin{figure}[hbt!]
\begingroup
    \centering
    \begin{tabular}{c c}
\hspace{-0.3cm}{ \includegraphics[width=0.45\linewidth]{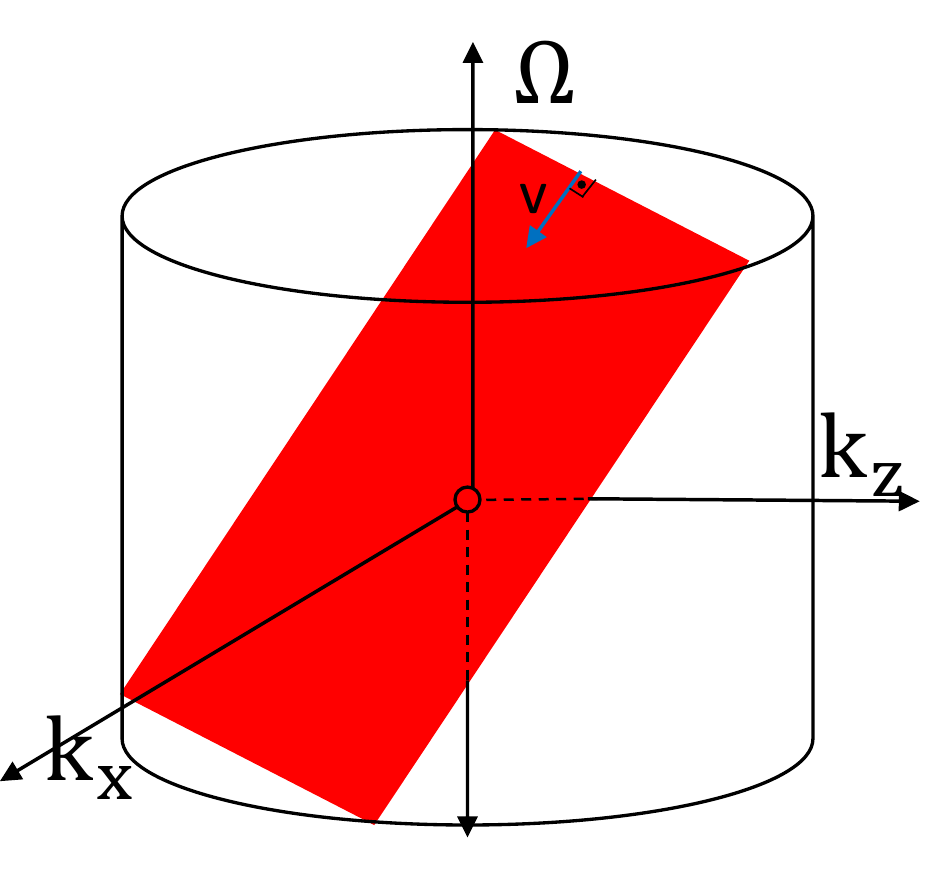}} & \hspace{-0.4cm}{\includegraphics[width=0.45\linewidth]{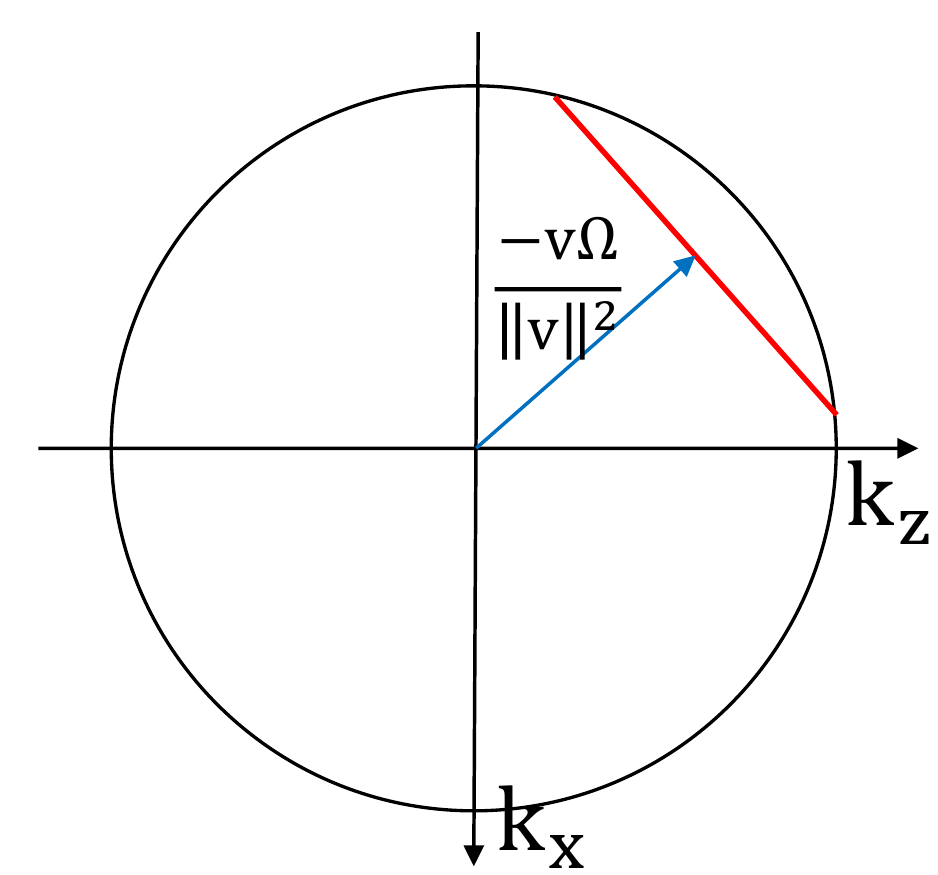}} \\
\end{tabular}
\caption{$B(\vc{k},\Omega )$ is supported on a plane in $(\vc{k},\Omega)$ space.}
\label{fig:planefig}
\endgroup
\end{figure}

The intuition for the approach is provided by investigating the 3D Fourier transform $B(\vc{k},\Omega)$ of the 3D ULM data \eqref{bubblesignal}. Under the assumptions, the signal $B(\vc{k},\Omega)$ due to one bubble moving at constant velocity $\vv$ and initial position $\vc{r_0}$ is found to be (Appendix A.1)
\begin{equation}
\label{eq:BkO}
    B(\vc{k},\Omega) = G(\vc{k})\exp(-i \vc{k}\cdot\vc{r_0}) \delta (\Omega + \vc{k}\cdot\vc{v})
\end{equation}
Due to the factor $\delta (\Omega + \vc{k}\cdot\vc{v})$, $B(\vc{k},\Omega )$ is nonzero only when $\Omega = - \vc{k}\cdot\vc{v}$. In other words, $B(\vc{k},\Omega )$ is a singular distribution supported on the plane  defined by $\Omega = - \vc{k}\cdot\vc{v}$ (Fig.~\ref{fig:planefig}). It follows that  signals due to microbubbles whose velocities are constant and identical, reside on a single common plane in spatio-temporal frequency space $(\vc{k},\Omega)$; different planes for different velocities. Hence,  to recover the microbubbles that are moving at a selected velocity $\vc{v_f}$, one should apply a filter whose
frequency response 
is supported on the plane  defined by $\Omega = - \vc{k}\cdot\vc{v_f}$, i.e., is equal to $\delta (\Omega + \vc{k}\cdot\vc{v_f})$.

In practice, to closely approximate the assumption of constant velocity motion, we window the spatiotemporal ULM data $b(\vc{r},t)$ 
by a short duration symmetric temporal window $w(t)$ of unit mass,  i.e. $w(t) = w(- t)$, and $\int_{-\infty}^\infty w(t) dt =1$. By centering the window at different times $t$, time varying velocities can be handled. The shorter the window, the greater the tolerance to deviation from constant velocity, but,
as will be shown, the velocity filter is more selective for longer windows. For example,  a reasonable choice for $w(t)$ is    
\begin{equation}\label{eq:w}
    w(t)  = 1/(\sqrt{2 \pi } \sigma_t) \exp\left(- t^2/(2\sigma_t^2)\right)
\end{equation}  
with the width parameter ${\sigma_t}$ controlling the trade-off between the constant velocity  assumption and velocity selectivity.

Because multiplication by the window in the time domain corresponds to convolution with respect to $\Omega$ in the frequency domain, the combined effect of  windowing and the ideal velocity filtering 
results in the velocity filter frequency response
\begin{align}
  H(\vc{k}, \Omega) & =   \delta (\Omega + \vc{k}\cdot\vc{v_f}) \ast_{\Omega} W(\Omega) \nonumber \\
  &= W(\Omega + \vc{k}\cdot\vc{v_f}) \label{eq:H}
\end{align}
This frequency response is again concentrated around the plane  $\Omega = - \vc{k}\cdot\vc{v}_f$, but
with non-zero width.  It 
is readily applied 
to 
the ULM data $b(\vc{r},t)$ using FFTs.

\subsection{Theoretical Analysis}
We analyze the effect of velocity filtering on the ULM data first for a generic PSF, then for pre-envelope PSF, and finally for post-envelope  PSF.  We obtain simple results describing the attenuation and distortion produced by the filter for microbubles moving at non-selected velocities. 
\subsubsection{Generic Case}
\label{sec:generic}
Multiplying \eqref{eq:BkO} and \eqref{eq:H} and using the identity $\delta (\Omega + \vc{k}\cdot\vc{v})W(\Omega+\vc{k} \cdot \vc{v_f}) = \delta (\Omega + \vc{k}\cdot\vc{v}) W(\vc{k} \cdot \vc{\Delta v})$ where $\vc{\Delta v} \triangleq \vc{v} - \vc{v_f}$, 
it follows that the
Fourier transform of the filtered ULM data is given by
\begin{equation}
    \Phi(\vc{k},\Omega)  = B(\vc{k},\Omega)H(\vc{k},\Omega)  
    = B(\vc{k},\Omega) W(\vc{k} \cdot \vc{\Delta v })
    \label{eq:PhikO}
\end{equation}
%
    \label{phi_kt_space2}
Note that  $\vc{\Delta v }$
is the difference between the known selected velocity and the actual bubble velocity, which is unknown. In other words, the signal $B(\vc{k},\Omega)$ corresponding to a buble moving at velocity $\vc{v}$ is filtered by the \emph{effective} (virtual) filter $\hat{H}(\vc{k} , \vc{\Delta v }) \triangleq W(\vc{k} \cdot \vc{\Delta v }) $. This  filter depends on the unknown \emph{actual} velocity $\vv$ of the microbubble, and therefore cannot be implemented; it only represents the effects of the actual filtering by the realizable filter $H$ in \eqref{eq:H}.

Because the effective filtering by $\hat{H}(\vc{k} , \vc{\Delta v })$ is only in spatial frequency, it corresponds to convolution $\ast_r$  in the space domain, 
\begin{align}
\label{rconvolvewitheffective}
   \phi(\vc{r},t) =  b(\vc{r},t) \ast_r \hat{h}(\vc{r} ,\vc{\Delta v }) 
\end{align}
Denoting by $\hat{\delv}$ and $\hat{\delv}^\perp$  unit vectors in the direction of $\delv$, and 
perpendicular to it, respectively, the impulse response  $\hat{h}(\vc{r} ,\vc{\Delta v })$ of the effective velocity filter is (Appendix A.2)
\begin{equation} \label{eq:hat-h}
\hat{h}(\vc{r} ,\vc{\Delta v }) = \begin{cases} \frac{1}{|| \vc{\Delta v}||} w\left(\frac{\hat{ \vc{\Delta v}} \cdot \vc{r} }{|| \vc{\Delta v}|| } \right) \delta(\hat{ \vc{\Delta v}}^\perp \cdot \vc{r} ) &  \vc{\Delta v} \not = 0\\ \delta(\vc{r} )& \vc{\Delta v} =0 \end{cases}  
\end{equation}
Equations \eqref{rconvolvewitheffective} and \eqref{eq:hat-h} reveal that the velocity selective filter enjoys the desired properties: 
\begin{itemize}
    \item[(P1)] It leaves the moving bubble untouched when $\delv=0$. 
    \item[(P2)] It provides attenuation and distortion when $\delv \neq 0$, but because the effective filtering is only in the spatial coordinates and not in time, it does not affect the motion of the filtered bubble.
    \item[(P3)] As $\|\delv\|$ increases, the attenuation and distortion increase. As a side note, the distortion has the form of stretching the microbubble response in the direction of $\delv$, leaving it unmodified in the orthogonal direction. 
\end{itemize}
\subsubsection{Pre-envelope Implementation} \label{basebandcase}
We now analyze the velocity filter applied to the beamformed image 
before envelope detection. We assume that the PSF can be approximated by the cosine-modulated Gaussian envelope 
\eqref{passbandPSF}, and use the Gaussian time window \eqref{eq:w}. To focus on the effect of the velocity filter on the magnitude and shape of the moving microbubble signal, we define its motion-free version 
\begin{align}
\label{motionfreebubble}
    q(\vc{r}, \vc{\Delta v})  \triangleq \phi(\vc{r}- \vc{s(t)},t) ,
\end{align} 
where $\vc{s(t)} = \vc{r_0}+\vc{v}t$, with the corresponding frequency domain relation
\begin{equation}
    Q(\vc{k},\vc{\Delta v }) = G(\vc{k}) W(\vc{k} \cdot \vc{\Delta v }) .
    \label{eq:QkDelv}
\end{equation}

Then, it is shown in Appendix A.3 that 
\begin{align}  \label{eq:q bandpass}
q(\vc{r},\vc{\Delta v})=& \Gamma(\vc{\Delta v}) g_e\left(\eta\vc{r}\right) \cos \left[ (2\pi/\lambda) \left( z- \zeta( \vc{r},\vc{\Delta v}) \right) \right] 
\end{align}
where 
\begin{align}
\label{gammaforpreenv}
    \Gamma(\vc{\Delta v}) &= \frac{1}{\sqrt{1 + \kappa^2 }} \exp\left[- \frac{2\pi^2}{1+ \kappa^2} \left(\frac{  \sigma_t {\Delta v}_z}{\lambda }
\right)^2 \right]    \\
\zeta(\vc{r},\vc{\Delta v}) &=   \kappa/(1+\kappa^2) (\sigma_t/\sigma_r) \cos \left(\theta_{\vc{r}, \vc{\Delta v}} \right) \|\vc{r}\| {\Delta v}_z \\
\label{etafactor}
    \eta &= \sqrt{
    1-\kappa^2/(1+\kappa^2) \cos^2(\theta_{\vc{r}, \vc{\Delta v} })
                 }\\
\label{kappafactor}
   \kappa &=(\sigma_t/\sigma_r) \| \vc{\Delta v}\|
\end{align}
and $\theta_{\vc{r}, \vc{\Delta v}}$ is the angle between $\vc{r}$ and $\vc{\Delta v}$.

It follows that for $\delv \neq 0$, the filtered bubble response is attenuated by factor $\Gamma(\vc{\Delta v})$, and distorted by factor $\eta(\vc{\Delta v}, \vc{r})$.
The phase of the modulation is also shifted by $\zeta(\vc{r},\delv)$. 
Here are some additional observations:
\begin{densitemize}
    \item The ratio $\sigma_r/\sigma_t$ is the speed $v_{r/t}$ for traveling a distance equal to width of the PSF in the duration of the time window. Hence $\kappa$ can be interpreted as a normalized version $\| \vc{\Delta v} \| /v_{b/w}$ of the magnitude of the velocity difference .
    
   \item For non-zero velocity difference component $\Delta v_z$ in the axial direction, the attenuation factor $\Gamma(\vc{\Delta v})$ has an exponential decay, which will provide rapidly increasing suppression with increasing $|\Delta v_z|$. Additional attenuation in this scenario is also due to the phase shift of the modulation, because the peak of the cosine no longer coincides with the peak of the Gaussian envelope. 
   Otherwise, the attenuation will be  by  factor $\approx 1/\kappa$, for $\kappa \gg 1$.
    
    \item The distortion factor $ 0 \leq \eta \leq 1 $ is a monotone decreasing function of both $\kappa$ and $|\cos(\theta_{r, \Delta V})|$, and determines how much the bubble response will widen: as $\eta$ approaches 1, the bubble response approaches the unfiltered case; as $\eta $ decreases,  the filtered microbubble response widens.
\item For $\vc{r} \parallel \vc{\Delta v}$, the distortion is maximized, with $\eta = 1 / \sqrt{1+\kappa^2}$. For $\kappa \gg 1$, this implies stretching of the velocity-filtered PSF by factor $\approx \kappa$ in the direction of $\vc{\Delta v}$.
\item For $\vc{r} \perp \vc{\Delta v}$, $\eta =1 $. This implies that the PSF width in the direction perpendicular to $\vc{\Delta v}$ is not affected by the filtering, and there is only attenuation in this direction.
\item The phase and amplitude distortions 
will lead to additional desirable attenuation of the output of the matched filter in response to ``rejected microbubbles'' in the localization step.
    \item $\sigma_t$ controls the trade-off between the constant velocity assumption and velocity selectivity. Increasing $\sigma_t$ provides stronger attenuation while imposing a stronger constant velocity assumption.
\end{densitemize}

\begin{figure}[hbt!]
\begingroup
    \centering
    \begin{tabular}{c c}
\hspace{-0.3cm}{ \includegraphics[width=0.5\linewidth]{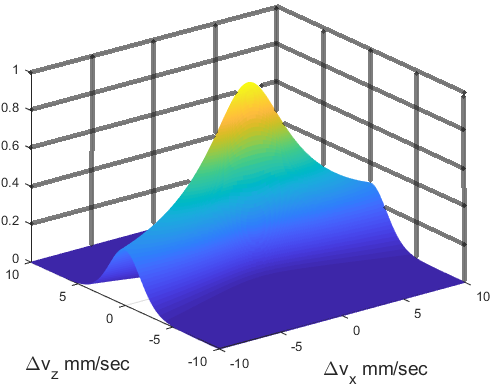}} & \hspace{-0.4cm}{\includegraphics[width=0.5\linewidth]{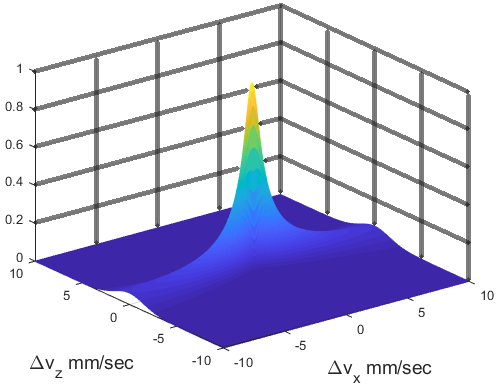} } \\
\end{tabular}
\caption{Attenuation factor $\Gamma(\vc{\Delta v})$ 
for $\sigma_r / \sigma_t = 3$mm/sec (left) and $\sigma_r / \sigma_t = 0.6$mm/sec (right).
}
\label{fig:Gamma_preenvelope}
\endgroup
\end{figure}

The attenuation factor $\Gamma(\vc{\Delta v})$ is plotted in Fig.~\ref{fig:Gamma_preenvelope}
The predicted strong effect of ${\Delta v}_z$ on the attenuation is evident. It is harder to achieve the same level of attenuation by the ${\Delta v}_x$ component only.


\subsubsection{Post Envelope Implementation} \label{passbandcase}
A similar analysis in the case of post-envelope detection velocity filtering expressions produces expressions similar to  \eqref{eq:q bandpass} -- \eqref{kappafactor}, except that the cosine factor in \eqref{eq:q bandpass} and the exponential factor in \eqref{gammaforpreenv} are missing. It follows that the attenuation and distortion for pre-envelope processing is much superior to that with post-envelope processing, and therefore pre-envelope velocity filtering is our method of choice.

\subsection{The Filter's Effect on Apparent Microbubble Density}
\label{secapparentmicrobubblechapter4}
\begin{figure}[hbt!]
\begingroup
    \centering
    \begin{tabular}{c}
\hspace{-1cm}{ \includegraphics[width=0.45\linewidth]{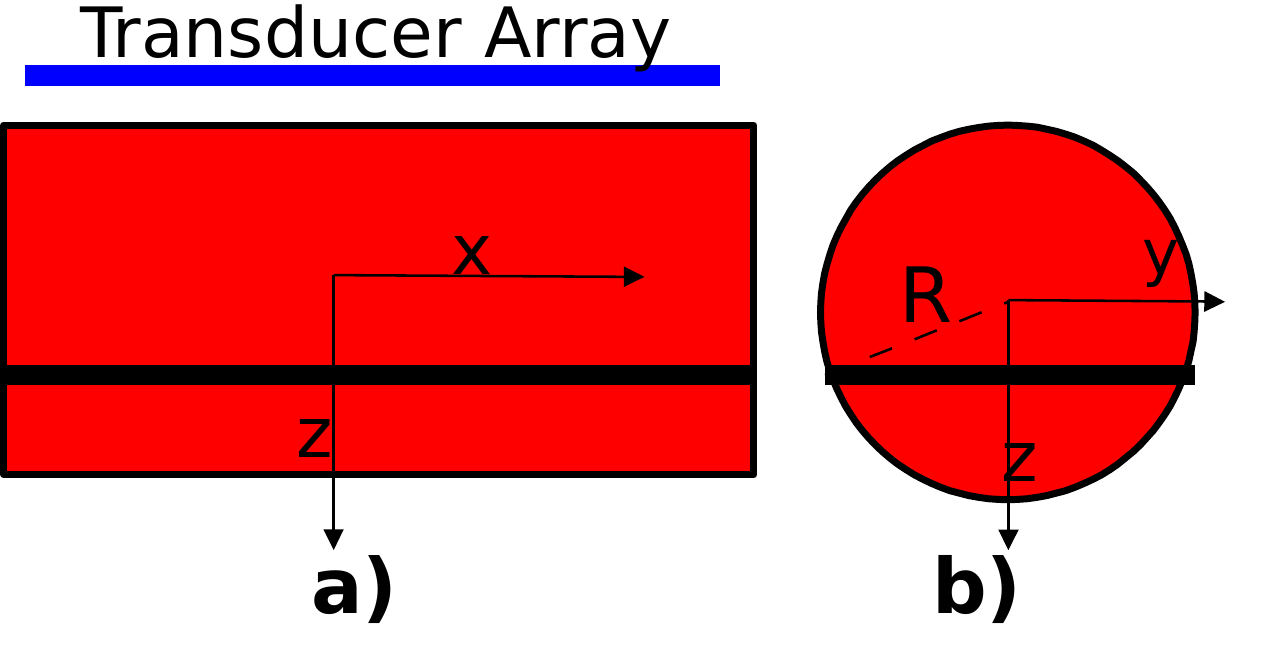}}
\end{tabular}
\caption{3D flow model: (a) XZ image plane (b) YZ Elevation-Depth plane.}
\label{3dflow}
\endgroup
\end{figure}

We examine now the effect of the velocity filter on the apparent microbubble density, using 
the 3D blood flow model of Sec.~\ref{bloodflowmodel}. Consider the single vessel with lateral flow in Figure \ref{3dflow}. In this subsection, for simplicity,  the directions of the actual microbubble velocity $\vc{v}$ and the velocity $\vc{v_f}$ selected by the velocity filter are assumed to coincide, and their magnitudes (corresponding speeds) are represented by $v$ and $v_f$, respectively.  The analysis can be generalized to $\vc{v}$ and $\vc{v_f}$ with different directions
by replacing $|v -v_f |$ by $\| \vc{\Delta v}\|$.
%

Because  microbubbles in the 3D vessel are projected in the beamformed image onto the 2D $(x,z)$ image plane, their density on this plane, which we call the \emph{apparent density}, will no longer be uniform.   Microbubbles located along the thick line in Fig.~\ref{3dflow}b, will be projected to same location on $(x,z)$ image plane. Likewise, those located on the plane parallel to the $(x-y)$ plane and cutting the vessel crosssection along the thick line in Fig.~\ref{3dflow}b, will be projected onto the thick line in Fig.~\ref{3dflow}a. Assuming the microbubbles have uniform concentration $C_{MB}$ throughout the vessel, the microbubble 2D apparent density is given by the line integral projection of the uniform density in the cylinder along $y$ onto the $(x,z)$ image plane. Now, note that the direction of the vessel in the image plane does not matter. Therefore,  we replace $z$ by the signed distance $\rho$ from the vessel center line, obtaining 
\begin{align}
\label{densitymodel}
    d_2(\rho) = 2 C_{MB} \sqrt{ R^2 - \rho^2}
\end{align}
As expected, the apparent density is independent of the coordinate along the vessel, but it does depend strongly on the distance from the vessel centerline. 

Another effect of the 3D to 2D projection in the beamformed image, is that microbubbles with different speeds located on the thick black line in Fig \ref{3dflow}b are projected to same location on the $(x,z)$ image plane, producing a distribution of speed values at each image location: different numbers of bubbles will have different speeds. Again, the relevant parameter is the distance $\rho$ from the vessel center line. We denote the joint microbubble density and speed distribution   by $d_2(v,\rho)$.  It  represents the density at distance $\rho$ from the vessel center of microbubbles that have speed in the differential interval $[v, v+dv]$, 
and has units of number of microbubbles per unit area and per unit speed. 

It is convenient to express $d_2(v,\rho)$ in terms of
  the speed  $v_0$ at the vessel center and the maximum speed $v_{\max}(\rho)$ at position $\rho$. By \eqref{parabolicformualspeed} we have
\begin{align}
\label{eq:vmax}
    v_{\max}(\rho) = v_0\left(1 - \rho^2/R^2\right)
\end{align}
It then follows (Appendix B) that 
\begin{align}
\label{d2z,vref}
 d_2(v,\rho) =& (C_{MB} R/v_{0}) [(1 - v/v_{\max}(\rho)) (1-(\rho/R)^2)]^{-\frac{1}{2}}   \nonumber \\& \times \rect[v/(2v_{\max}(\rho)) ] \rect[\rho/(2R)] 
\end{align}




Our goal is to determine the effect of the velocity filter on the apparent microbubble density $d_2(\rho)$. To this end, we first determine the velocity range (\emph{velocity pass-band}) that is passed by the velocity filter. This must account for the effects of both the velocity filter and the bubble detection/rejection in the localization step. We have already analyzed the first. For the second, recall  that our model (Sec.~\ref{basiclocamethod}) for localization by matched filtering assumes that the matched filter output is thresholded to detect bubbles. We proceed with the analysis by making the reasonable assumption that the threshold level is chosen as $1/2$ of the autocorrelation peak. It follows that the condition for rejecting a filtered microbubble, and on $ \vc{\Delta v}$ to be outside the velocity pass-band is:
\begin{align}
\label{rangeofvelocities}
    q(\vc{r}, \vc{\Delta v}) \ast g(-\vc{r}) < 
    0.5\max_{\vc{r}'} g(\vc{r}') \ast g(-\vc{r}')   \text{ }\forall \vc{r} \in \mathbb{R}^2
\end{align}
where $g(\vc{r})$ is the PSF and $q(\vc{r}, \vc{\Delta v})$ is the motion-free filtered microbubble signal. Subject to the assumption that $v$ and $v_f$ are co-linear, the pass-band will be the interval $R(v_f) =[v_f-\delta v ,v_f+ \delta v] $
 around $v_f$,
where the velocity bandwidth $\delta v$ is determined as the smallest value of $\|\vc{\Delta v} \| = | v- v_f|$ for which \eqref{rangeofvelocities} is satisfied.

\begin{figure}[hbt!]
\begingroup
    \centering
\hspace{0cm}{ \includegraphics[width=0.7\linewidth]{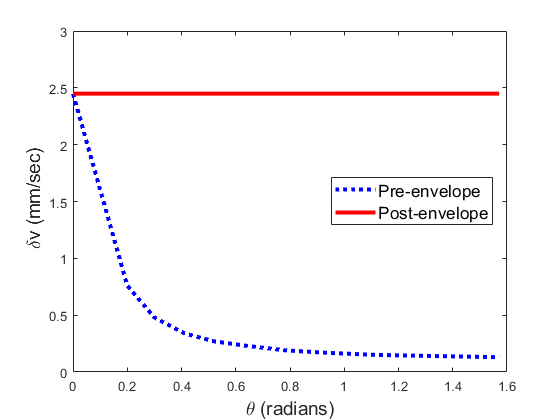}} \\
\caption{
$\delta v$ for $\sigma_r / \sigma_t =1 mm/sec$ as a function of the angle between $\vc{\Delta v}$ and the lateral axis.}
\label{deltavfig}
\endgroup
\end{figure}

The details of determining $\delta v$ can be found in Appendix D. 
It turns out that $\delta v $ is proportional to $\sigma_r / \sigma_t$, consistent with the expectation that the velocity filter is more selective for larger $\sigma_t$, but $\delta v $ also depends nonlinearly on the direction of $\vc{\Delta v}$. This is seen in Fig.~\ref{deltavfig}, where $\delta v$ is plotted versus the angle between $\vc{\Delta v}$ and the lateral axis.
Whereas $\delta v$ is independent of direction in the case of post-envelope velocity filtering, for pre-envelope velocity filtering, $\delta v$ decreases, and the velocity filter becomes more selective, 
 as the direction of $\vc{\Delta v}$ gets closer to axial. 


Let $d_{VF}(\rho; v_f)$ denote the apparent microbubble density when the velocity filter is tuned to $v_f$; 
it is found (Appendix B) 
by integrating $d_2(v,\rho)$ with respect to $v$ over the velocity pass-band $R(v_f)$ defined earlier, yielding 
\begin{align}
\frac{d_{VF}(v_f,\rho) }{2 C_{MB} R} &= \sum_{j=0}^1 \sqrt{\left( 1 - \frac{v_f - (-1)^j\delta v}{v_{0}(\rho)} \right)\left(1 -\left(\frac{\rho}{R}\right)^2 \right)} \nonumber \\
& \times \rect \left(\frac{v_f - (-1)^j\delta v }{v_0(\rho)} -\frac{1}{2}\right)
\rect\left(\frac{\rho }{2R}\right)
\end{align}

\begin{figure}[hbt!]
\begingroup
    \centering
    \begin{tabular}{c c}
\hspace{-0.7cm}{ \includegraphics[width=0.42\linewidth]{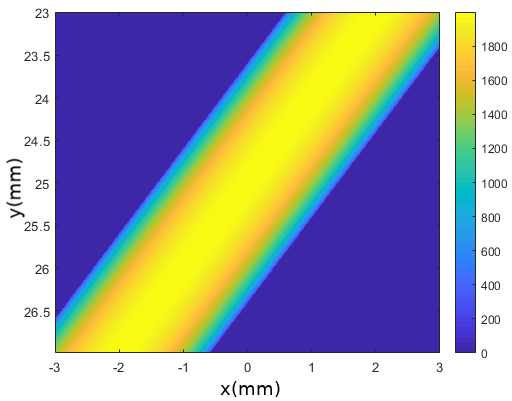}} & 
\hspace{-0.3cm}{\includegraphics[width=0.42\linewidth]{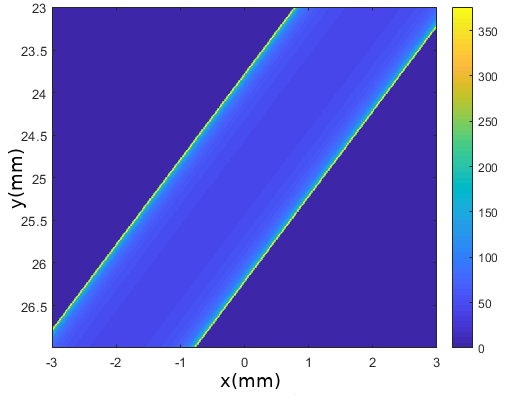}}
\end{tabular}
\caption{Velocity filter's effect on apparent microbubble density for diagonal flow:  $d_2(z)$ left,  and $d_{VF}(v_f,z)$ right, for $v_f = 2.5$ mm/sec, $v_{0}= 10$ mm/sec, $R = 1$ mm,  $\sigma_r /\sigma_t =1$ mm/sec and $C_{MB}= 10^3$ MBs/mm$^3$. The colorbars represent number of microbubbles per mm$^2$. (Note the different scales on left and right).} 
\label{apparentfig1}
\endgroup
\end{figure}

\begin{figure}[hbt!]
\begingroup
    \centering
    \begin{tabular}{c c c c}
\hspace{-0.5cm}{ \includegraphics[width=0.28\linewidth]{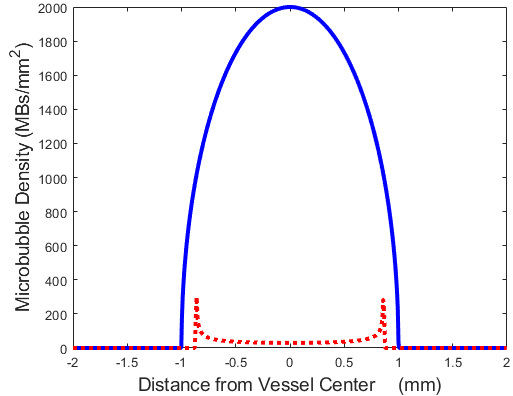}} & \hspace{-0.5cm}{ \includegraphics[width=0.24\linewidth]{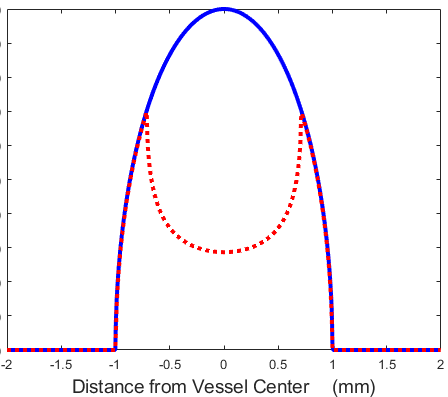}}&\hspace{-0.5cm}{ \includegraphics[width=0.24\linewidth]{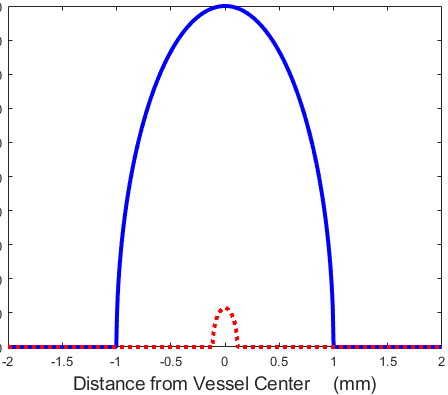}} & \hspace{-0.5cm}{ \includegraphics[width=0.24\linewidth]{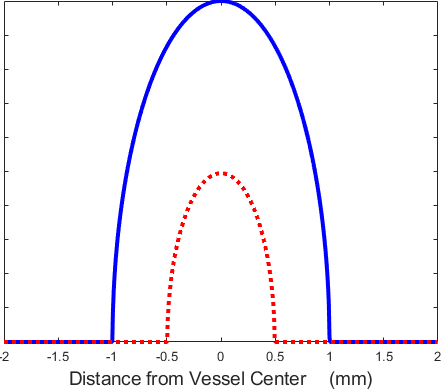}} \\
\hspace{-0.5cm}{\footnotesize (a) }&{\footnotesize \hspace{-0.5cm}(b)}&{\footnotesize \hspace{-0.5cm}(c) }&{\footnotesize \hspace{-0.5cm}(d) } \\
\end{tabular}
\caption{
Velocity filter's effect on the apparent microbubble density for $R = 1$ mm, $v_{0}= 10$ mm/sec, $\sigma_r /\sigma_t =1$ mm/sec and $C_{MB}= 10^3$ MBs/mm$^3$: (a) Axial $\Delta v$, $v_f = 2.5$ mm/sec  (b) Lateral $\Delta v$, $v_f = 2.5$ mm/sec (c) Axial $\Delta v$, $v_f = 10$ mm/sec (d) Lateral $\Delta v$, $v_f = 10$ mm/sec. Solid (blue) line: represents $d_2(\rho)$;  Dotted (red) line: $d_{VF}(\rho;v_f)$. } 
\label{apparentfig2}
\endgroup
\end{figure}

To visualize the effect of the velocity filter on the apparent microbubble density, Fig.~\ref{apparentfig1} depicts  $d_2(\rho)$ and $d_{VF}(\rho;v_f)$  for diagonal flow. It can be seen that the velocity filter reduces the apparent microbubble density dramatically. This is further explored for axial and lateral flow by plotting $d_2(\rho)$ and $d_{VF}(\rho;v_f)$ in Fig.~\ref{apparentfig2}, corresponding to slicing the apparent density perpendicular to blood flow.  Fig.~\ref{apparentfig2} further confirms that the velocity filter  reduces the apparent microbubble density significantly. Moreover, when $v_f$ is equal to the speed at the center line of the vessel, the apparent microbubble density vanishes around vessel boundaries. This characteristic of the velocity filter can be used to identify closely located blood vessels as separate vessels.

A final observation from this analysis is that the velocity filter provides smaller reduction of the apparent bubble density for lateral than for axial flow. This limitation is addressed in the next subsection by transverse oscillations. 

\subsection{Transverse Oscillation}
We have seen in Secs.~\ref{passbandcase} and \ref{secapparentmicrobubblechapter4} much stronger velocity selectivity in the axial than in the lateral direction. This is due to the ${\Delta v}_z$-dependent exponential decay factor in the attenuation factor $\Gamma(\vc{\Delta v})$;  with only a nonzero ${\Delta v}_x$ component, the velocity filter produces a much weaker inverse linear attenuation  $ \Gamma((\Delta v_x,0)) \propto
\approx 1/\Delta v_x$ for $\kappa>>1$. 
This difference between the axial and lateral responses is due to the anisotropy of the PSF:   
the exponential decay of $\Gamma$  with ${\Delta v}_z$  arises thanks to  the cosine modulation of  $g(\vc{r})$ in the axial direction. With the goal of obtaining an exponential decay of $\Gamma$  with lateral velocity difference ${\Delta v}_x$ too, 
 we analyze in this section the effects of introducing a similar modulation of $g(\vc{r})$ in the lateral direction. 
\subsubsection{Transverse Oscillation -- Implementation}
\label{TOImlemet}
The idea of introducing lateral (or transverse) oscillation into the PSF of the beamformed ultrasound image has been studied before in the context of doppler-based flow imaging, 
where it is known as transverse oscillations (TO). 
Whereas classical TO is implemented by means of
 a special apodization function in the receive beamforming step, 
Varray and Liebgott \cite{varray2013alternative} propose to process the beamformed image by filtering along the lateral coordinate. Their proposed  
TO filter selects k-space components around a desired lateral oscillation frequency and has the frequency response 
\begin{align}
\label{TOfilterspec}
    G_T\left(\vc{k}\right)&=\exp\left[
    -\sigma_{x}^2\left(k_{x}-k_{0 x}\right)^2/2 \right] \nonumber \\
    &  \, +\exp\left[-\sigma_{x}^2\left(k_{x}+k_{0 x}\right)^2/2\right] 
\end{align} 
where 
$\sigma_x$ controls the lateral width of the envelope of the resulting PSF.

Using our model \eqref{basebandPSF}-\eqref{passbandPSF} for the beamformed PSF, we find (Appendix E.1) 
that after filtering by the TO filter the PSF becomes
\begin{align}
\label{TOPSF}
g^{TO}(\vc{r})  =  g^{TO}_e(\vc{r}) \cos\left(  2 \pi z/ \lambda_z \right) \cos\left(  2 \pi x/ \tilde{\lambda}_x \right)
\end{align}
where $\lambda_x =2 \pi/k_{0x}$ and $ \tilde{\lambda}_x \defsign (1+\sigma_r^2/\sigma_x^2) \lambda_x$ is the lateral oscillation wavelength and the envelope of the PSF is
\begin{align}
    \label{eq:geTO}
     g^{TO}_e(\vc{r}) &= \frac{2}{ 
     \sqrt{(1 +\sigma_x^2/\sigma_r^2)}}
\exp\left[
-\frac{ 2 \pi^2 \sigma_r^2}{\tilde{\lambda}_x^2} \left(1+\frac{\sigma_r^2}{\sigma_x^2}\right)
\right] \nonumber \\
& \quad \times g_e\left(\diag [ (1+\sigma_x^2/\sigma_r^2)^{-1/2}, 1 ]
\vc{r} \right) .
\end{align}
Thus, the TO filtering introduces the desired lateral modulation into the PSF, along with a modified envelope.

\subsubsection{Theoretical Analysis of TO}
\label{TOcasetheory}
It is shown in Appendix E.2 
that in the presence of TO the motion-free filtered microbubble signal is
\begin{align}
\label{TOtheory}
 q(\vc{\Delta v},\vc{r}) = g^{TO}_e\left[\Xi(\vc{\Delta v}) \vc{r} \right]\left(\Gamma_1 \cos \theta_1 +\Gamma_2 \cos \theta_2  \right)
\end{align}
where for $j=1,2$
\begin{align*}
\Gamma_j &= (2 \sqrt{1 + \tilde{\kappa}^2 })^{-1} e^{- \frac{2\pi^2 \sigma_t^2}{1+ \tilde{\kappa}^2} \left(\frac{   {\Delta v}_z}{\lambda_z } - \frac{ (-1)^j    \Delta v_x}{\tilde{\lambda_x} } \right)^2} \\
\theta_j &=2\pi z/ \lambda_z - (-1)^j 2\pi  x/ \tilde{\lambda_x}  \\
& -\left(\frac{2\pi \sigma_{t}^2}{1 + \tilde{\kappa}^2}\right)  
\left( \frac{{\Delta v}_z}{\lambda_z} - \frac{(-1)^j {\Delta v}_x}{\tilde{\lambda_x}} \right)
\vc{\Delta v}^T D \vc{r} \\
\tilde{\kappa}^2 &= \sigma_t^2 \vc{\Delta v}^T D \vc{\Delta v}^T\\
%
\Xi(\vc{\Delta v}) &= I - \frac{\sigma_t^2}{\tilde{\kappa}^2} \left(1- \frac{1}{\sqrt{1+\tilde{\kappa}^2}}
    \right)\vc{\Delta v} \vc{\Delta v}^T D
\end{align*}
and $D= \diag \left[ (\sigma_r^2+\sigma_x^2)^{-1} , \sigma_r^{-2}
    \right]$.
The peak of $q(\vc{\Delta v},\vc{r})$ occurs at $\vc{r}=0$, at which point $\theta_1=\theta_2=0$. Hence, the attenuation of the microbubble response by the velocity filter with TO can be judged by considering the upper bound
\begin{align}
\label{gammabar}
    {\footnotesize \Gamma_1 \cos \theta_1 + \Gamma_2 \cos \theta_2 \leq \Gamma_1 +\Gamma_2 \triangleq \overline{\Gamma}}
\end{align}
Furthermore, 
consider the following special cases of \eqref{TOtheory}.\\
\textbf{Case (i) ${\Delta v}_z = 0$:} \vspace*{-0.4cm}
\begin{equation*}
     q(\vc{r},\vc{\Delta v}) =  \Gamma_x(\vc{\Delta v}) g^{TO}_e\left(\Xi_x \vc{r} \right) \cos\left(\frac{2\pi z}{\lambda_z}  \right) \cos \left(\frac{2\pi x}{\tilde{\lambda_x}}- \zeta_x \right)
\end{equation*}
\textbf{Case (ii) ${\Delta v}_x = 0$:} \vspace*{-0.4cm}
\begin{equation*}
     q(\vc{r},\vc{\Delta v}) =  \Gamma_z(\vc{\Delta v}) g^{TO}_e\left(\Xi_z \vc{r} \right) \cos\left( \frac{2\pi z}{\lambda_z}- \zeta_z \right) \cos \left( \frac{2\pi x}{\tilde{\lambda_x}} \right)
\end{equation*}
where, for $j=x,z$ (Note: $\tilde{\lambda}_z = \lambda_z$)
\vspace*{-0.2cm}
\begin{align*}
    \Gamma_j(\vc{\Delta v}) &=(1 + \tilde{\kappa}^2 )^{-\frac{1}{2}} e^{- \frac{2\pi^2}{1+ \tilde{\kappa}^2} \left(\frac{  \sigma_t \Delta v_j}{\tilde{\lambda}_j } 
\right)^2} \\
\zeta_j &= 2 \pi j \frac{
\sigma_{t}^2 {\Delta v}_j^2
                      }
                      {
                      \tilde{\lambda}_j (1 + \tilde{\kappa}^2) 
                      (\sigma_r^2 + \mathbb{1}_{x=j}\sigma_x^2)
                      }
\end{align*}
\[\Xi_x=\Xi[({\Delta v}_x,0)] \qquad \Xi_z=\Xi[0, ({\Delta v}_z)]]
\]
\textbf{Case (iii) ${\Delta v}_z/\lambda_z = {\Delta v}_x/\tilde{\lambda_x} $:} For this case, $\Gamma_2 \gg \Gamma_1$ and \vspace*{-0.2cm}
\begin{equation*}
     q(\vc{r},\vc{\Delta v}) \approx   g^{TO}_e\left(\Xi \vc{r} \right) /( 2 \sqrt{1 + \tilde{\kappa}^2 }) \cos\left[ 2\pi \left( z/\lambda_z- x/\tilde{\lambda_x}   \right)\right]
\end{equation*}
\textbf{Case (iv) ${\Delta v}_z/\lambda_z = -{\Delta v}_x/\tilde{\lambda_x} $:} For this case, $\Gamma_1 \gg \Gamma_2$ and \vspace*{-0.2cm}
\begin{equation*}
     q(\vc{r},\vc{\Delta v}) \approx   g^{TO}_e\left(\Xi \vc{r} \right) /( 2 \sqrt{1 + \tilde{\kappa}^2 }) \cos\left[ 2\pi \left( z/\lambda_z+ x/\tilde{\lambda_x}   \right)\right]
\end{equation*}
Examining Cases (i) and (ii), we conclude that with TO, the velocity filter enjoys exponential decay with respect to $\vc{\Delta v}$ in both the lateral and transverse directions. However, as seen in Cases (iii) and (iv), two specific directions will lose the exponential decay in the attenuation. 

\begin{figure}[hbt!]
\begingroup
    \centering
    \begin{tabular}{c c}
\hspace{-0.3cm}{ \includegraphics[width=0.45\linewidth]{figures/preenvelopegamma/threed0point1new2.png}} & \hspace{-0.4cm}{ \includegraphics[width=0.45\linewidth]{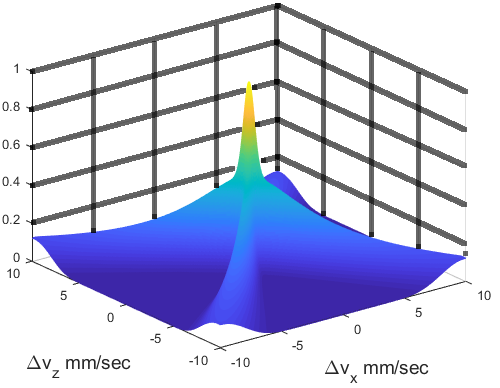}} \\
{\footnotesize
\hspace{-0.4cm} a) Velocity Filter without TO ($
\Gamma$)} & {\footnotesize \hspace{-0.7cm} b) Velocity Filter with TO($\overline{\Gamma}$)
}
\end{tabular}
\caption{Effect of Transverse Oscillation (TO) on microbubble attenuation by the velocity filter.
$\Gamma$ and $\overline{\Gamma}$ correspond to \eqref{eq:q bandpass} and \eqref{gammabar}, respective.
The lateral oscillation frequency for TO is chosen identical to the axial oscillation frequency.}
\label{TOvswithoutTO2}
\endgroup
\end{figure}

 To further study the effect of TO in the velocity filter on the anisotropy of the attenuation factor, we plot $\Gamma$  \eqref{eq:q bandpass} and $\overline{\Gamma}$  \eqref{gammabar}  in Fig.~\ref{TOvswithoutTO2}.
Fig.~\ref{TOvswithoutTO2} confirms that  without TO, $\Gamma$ decays slowly along the lateral direction. On the other hand,  with TO, the attenuation factor $\overline{\Gamma}$ has rapid decay in both the axial and lateral directions, but  decays slower along diagonal directions.   

In view of the foregoing analysis and Fig. \ref{TOvswithoutTO2}, we expect that including TO with velocity filtering will be useful for reconstructing vessels with flow close to the lateral direction.  There, for $\vc{v_f}$ chosen to select the vessel, $\vc{\Delta v}$ across the vessel will point in the lateral direction too, and strong selectivity with respect to $\Delta v_x$ will be important for improving the reduction of the apparent microbubble density from the scenarios in Fig.~\ref{apparentfig2}-b or Fig.~\ref{apparentfig2}-d to ones similar to Fig.~\ref{apparentfig2}-a and Fig.~\ref{apparentfig2}-c, respectively.
On the other hand, TO should be left out for diagonal flows. Hence, an overall pipeline could take advantage of TO with velocity filtering for resolving lateral flows, while leave it out for resolving other directions.    

\subsection{Effects on Noise and Penetration Depth}
\label{noiseanalysis}
The velocity filter offers increased noise rejection, and therefore increased penetration depth for ULM. We focus on electronic noise, because the time-invariant speckle noise can be rejected along with reflections from static tissue by existing techniques for removal of the static image component \cite{demene2015spatiotemporal}, and would also be suppressed by the velocity filter itself.

We first determine passband in $(\vc{k}, \Omega)$ space of the standard processing chain. Let $k_G$ be the spatial bandwidth of the ultrasound system MTF. 
Then the spectrum $B(\vc{k}, \Omega)$ of moving microbubble responses is supported on the intersection of the plane and cyliner of radius $k_G$ in Fig.~\ref{fig:planefig}. Hence, it is bandlimited in temporal frequency to $\Omega_{\max} = k_G v_{0,\max}$, where $v_{0,\max}$ is the maximum flow speed in the imaged volume. 
Assuming that the electronic noise in each beamformed image frame has a flat power spectrum \cite{michailovich2006despeckling}  bandlimited to $\|\vc{k}\| \le k_G$, 
and using the statistical independence of the noise from frame to frame, we can employ standard results for the propagation of a wide-sense stationary process through a linear shift invariant system to perform the theoretical analysis.

Let $E[(n)^2(\vr,t)]$ be the power of the input noise to the
localization step without velocity filtering, and $E[(n')^2(\vr,t)]$ the noise power after velocity filtering, and define the noise reduction factor ($NRF$) as the ratio of the two. Then it is shown in Appendix F that 
\begin{equation*}
    NRF \defsign E[(n)^2(\vr,t)]/E[(n')^2(\vr,t)] \geq  (2/\sqrt{\pi}) k_G v_{0,\max} \sigma_t
\end{equation*}  

For instance, for $\sigma_t = 0.5$sec, $v_{0,\max}=10$ mm/sec, $\lambda =0.3$ mm, and $k_G=2\pi/\lambda$, we obtain $NRF = 118$ which corresponds to 21dB noise power reduction. 

To interpret, note that the  attenuation of ultrasound in muscle/brain tissue is  $\approx 0.5$ dB/cm/MHz \cite{book}. At center frequency of 5MHz, the attenuation is $\approx5$ dB/cm for the roundtrip. It follows that the 21dB noise reduction can increase  penetration depth by about 4cm, which just on its own could enable new upplications of ULM in medicine and biology.

\subsection{ULM Performance Metrics}
We aim to recover two features: vessel structure, and blood flow information,
and therefore propose 
corresponding 
metrics to evaluate the effectiveness of the proposed velocity filtering.

\subsubsection{Vessel Structure}
We propose two metrics: (i) Localization Error ($LE$); and (ii) Intersection Over Union ($IoU$).

First, we introduce the $LE$ metric, to assess the success of the algorithm in recovering an accurate map of the locations of microbubble centers in each of the beamformed image frames.
Given a ground truth image frame  $V(\vc{r})$ and a corresponding estimated image frame $\hat{V}(\vc{r})$, each with
points representing microbbuble positions, the goal is to asses the accuracy of $\hat{V}(\vc{r})$. Ideally, the accuracy metric should be proportional to e.g., the mean-squared position error of the points.

This raises several challenges. First, consider the standard squared error $\|\hat{V}(\vc{r}) - V(\vc{r})\|_2^2$ between the true and recovered images with point distributions. For a pair of points with true and estimated positions $\vc{r}_t$ and $\hat{\vc{r}}_{t}$, the squared error will be the same for all $\| \vc{r}_t - \hat{\vc{r}_{t}} \|_2$ except  when the points overlap, i.e., 
$\vc{r}_t = \hat{\vc{r}}_{t}$, and their contribution to the squared error vanishes. Hence, the standard MSE between the image frames cannot be used to quantify the localization error. Although it is possible to use standard methods for comparing point clouds,  these typically require special treatment when the number of points in the two clouds do not coincide. 

Instead, we introduce the following simple
metric, which is furthermore tailored to the ULM task:
\begin{align}
  \label{LEmetric}
  LE \triangleq 2(\sigma_\| \sigma_\perp \pi T)^{-1} ||[\hat{V(\vc{r})}  - V(\vc{r}) ] \ast_{\vc{r}} e(\vc{r})  ||_{2}^2
\end{align}
where $T$ is the true number of microbubbles in the scene,
and $e(\vc{r})$ is the following Gaussian blur kernel 
\begin{align}
\label{modifygauss}
    e(\vc{r}) = \exp\left[ -0.5 \vc{r}^T R(\theta)^T \Sigma^{-1} R(\theta) \vc{r}   \right]
\end{align}
where 
$ \Sigma = \begin{bmatrix} \sigma_\|^2 &0 \\ 0 &\sigma_\perp^2 \end{bmatrix} \quad R(\theta) = \begin{bmatrix} \cos\theta &-\sin\theta \\ \sin\theta & \cos\theta \end{bmatrix}
$.
The rotation angle $\theta\in [-\pi, \pi]$ of the rotation matrix $R(\theta)$ is chosen to coincide with the flow direction of the blood vessel to be reconstructed and selected by the velocity filter. Parameters $\sigma_{\|}$ and $\sigma_{\perp}$ control the width of the blur kernel parallel to perpendicular to the flow direction, respectively. 

The inclusion of the blur kernel makes $LE$  approximately proportional  to the mean-square 
 microbubble localization error. The particular anisotropic shape of the blur kernel makes $LE$ more sensitive to localization errors perpendicular than to those parallel to the flow direction, because the former will cause greater degradation in recovered vessel structure, possibly moving recovered microbubbles to outside vessel boundaries. 
To verify these claims,
consider a single microbubble in each of $\hat{V(\vc{r})}$ and $V(\vc{r})$. Let $\vc{r}_{t}$ and $\hat{\vc{r}_{t}}$ denote the true and recovered positions of the bubble, respectively. Then, using Taylor expansion and ignoring high-order terms, \eqref{LEmetric} becomes  
\begin{align}
\label{Taylorser}
  LE \approx (2/(\sigma_\| \sigma_\perp \pi T)) || \nabla_{\vc{r}}^Te(\vc{r} -\vc{r}_{t}) (\vc{r}_{t} -\hat{\vc{r}_{t}}) ||_{2}^2   
\end{align}
where $\nabla_{\vc{r}}e(\vc{r})$ is the gradient of $e(\vc{r})$ with respect to $\vc{r}$. 
Evaluating the 2-norm  \eqref{Taylorser} becomes
\begin{align}
\label{LEfinalexp}
    LE &\approx   \|A( \vc{r}_t - \hat{\vc{r}_{t}} )\|_2^2
\end{align}
 where $A \defsign \Sigma^{-1/2} R(\theta)$. Equation \eqref{LEfinalexp} suggests that the $LE$ metric and the actual localization error $\|A( \vc{r}_t - \hat{\vc{r}_{t}} )\|_2^2$, which is the weighted squared distance between recovered microbubble and true position microbubble positions, coincide to first order. 
 Furthermore, $LE$ automatically handles a mismatch between the number true and detected microbubble positions (e.g., due to missed or false detections), and is cheap to compute.

The second  structural recovery metric we use is the classical intersection over union ($IoU$). It measures the similarity between the  true vessel support and the one determined from the final output of the overall processing pipeline of Sect.~\ref{basiclocamethod} after semantic segmentation. It is given by
\begin{align}
\label{IoUdefinition}
  IoU = \operatorname{Area}(P_{c} \cap \hat{P_{c}})/ \operatorname{Area}(P_{c} \cup \hat{P_{c}})  
\end{align}
where $\hat{P_{c}}$ and $P_c$ is are the recovered and true vessel supports, respectively. We have $0 \leq IoU \leq 1$,  and larger $IoU$ values indicate better resemblance between $P_{c}$ and $\hat{P_{c}}$. 
\subsubsection{Blood Flow}
The algorithm can be used to produce a recovered pixelated velocity map $\hat{V}(\vc{r})$  with a  vector at each pixel showing the estimated velocity on the center plane of the recovered vessel. This map is obtained by  assigning the velocity of the fastest moving recovered microbubbles to any pixel position. Given a corresponding ground truth velocity map $V(\vc{r})$ we introduce the following flow velocity error ($FVE$) metric,  to evaluate recovered velocity map map:
\begin{align}
\label{FVEdef}
  FSE = \|V(\vc{r}) - \hat{V}(\vc{r}) \|_1 / P_v  
\end{align}
where $P_v$ is the number of pixels that have non-zero speed value in the ground truth velocity map. Thus, $FVE$ shows average velocity error per pixel.
\section{Numerical Experiments}
\label{numericalexpchapter}
\begin{figure}[hbt!]
\begingroup
    \centering
    \begin{tabular}{c }
\hspace{-0.3cm}{ \includegraphics[width=1.02\linewidth]{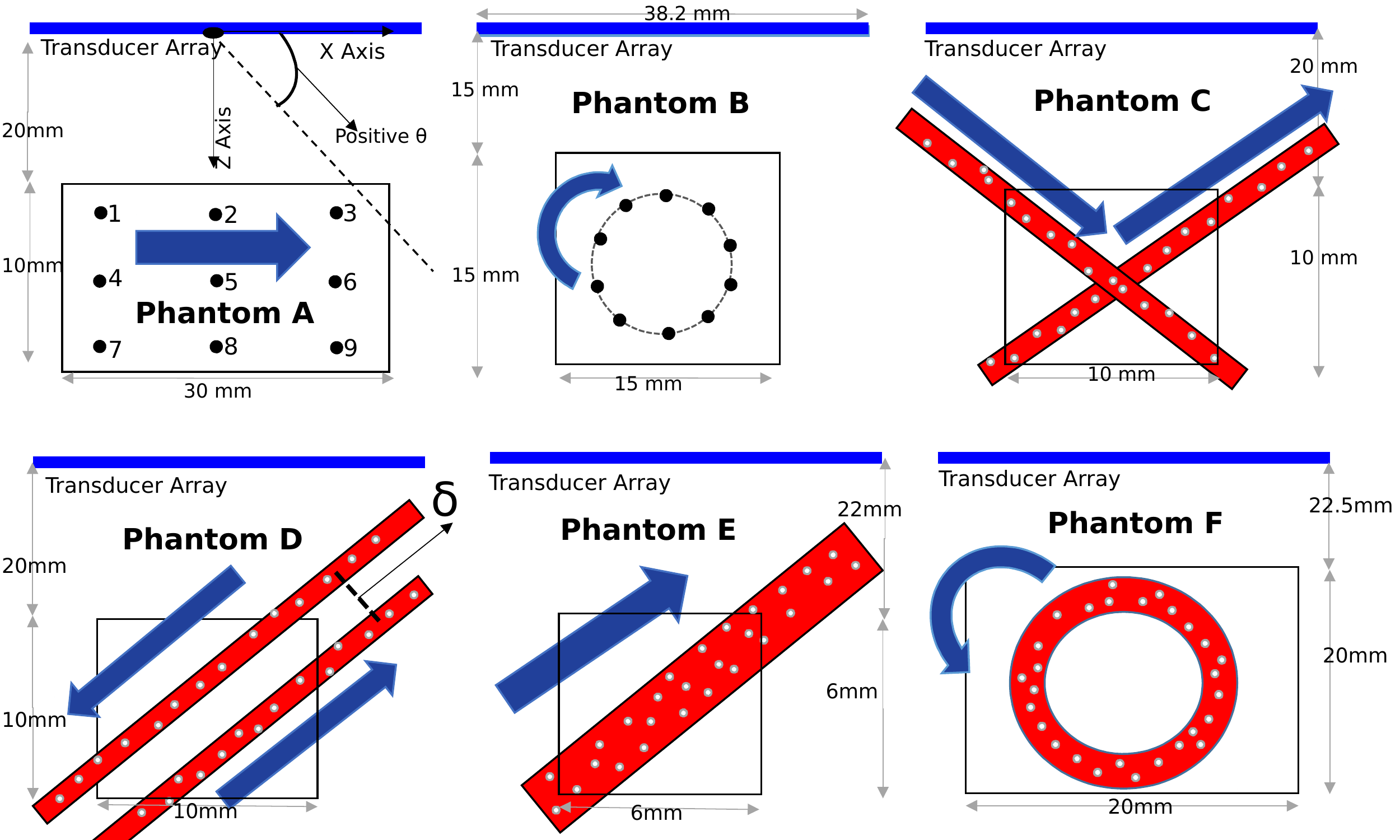}}\\
\end{tabular}
\caption{Simulation geometries.}
\label{Phantom Simulations VF}
\endgroup
\end{figure}

For a realistic ultrasound simulation, 
we employ the 3D flow model in Sec.~\ref{bloodflowmodel}, and (under Matlab R2018a) the widely used ultrasound simulation program Field II \cite{jensen1996field} .
A 1D linear array is used with coherent plane wave compounding.
We use for the study mathematical microvessel phantoms with geometries shown in Fig.~\ref{Phantom Simulations VF} and specifics in Table \ref{tab:my_label_spec_VF}, with microbubbles modeled as point scatters.
\begin{table}[htb!]
    \centering
  \caption{ \footnotesize simulation specifics }
{\footnotesize
\begin{tabular}{ |p{3.5cm}||p{4.4cm}|  }
\hline
Simulation Parameter &  Values/Expressions\\
 \hline
Transducer Element Height   & 5 millimeters  \\
Transducer Element Width&   0.27 millimeters  \\
Kerf & 0.03 millimeters\\
Number of Transducers    & 128\\
Center Frequency of Pulse & 5 MHz\\
Pulse Shape & Hanning windowed cosine signal\\
Pulse Duration & 2 Periods\\
Frame Rate& 100 Hz\\
Pixel Dimensions $D_X$, $D_Z$ & $\lambda/10$\\
Number of Plane Waves & 3 (-4,0,4 degrees)\\
Simulation Duration &3 seconds\\
Microbubble Speed & 1 mm/sec\\
Circular Motion Radius & 6.7 mm\\
 \hline
\end{tabular}
}
\label{tab:my_label_spec_VF}
\end{table}



\subsection{Numerical Experiments for Velocity Filtering}
\label{numexpforve}
In this subsection, we study the effects of the velocity filter using microbubble motion phantoms A and B in Fig.~\ref{Phantom Simulations VF}.

Phantom A has 9 microbubbles arranged in a square grid on a water background, all moving at the same speed of 1 mm/sec (typical in microvessels \cite{ivanov1981blood}) laterally, that is, with velocity $\vc{v} = 1 \angle{0^{\circ}}$ (positive angle is defined from the positive $x$ axis to the positive $z$ axis.) Their positions were chosen to bring out the effect of spatially varying PSFs.  

Phantom B has 10 microbubbles on a water background uniformly spaced on a circle of radius 6.7 mm and moving on it at 1 mm/sec. Because it violates the constant velocity assumption on which the velocity filtering approach is predicated, circular motion provides a critical test for the performance of the entire approach. It is especially interesting to investigate the trade-off between the constant velocity assumption and velocity selectivity controlled by $\sigma_t$. 
The centripetal acceleration (given by speed$^2$/radius) experienced by the microbubbles is $\sim0.15$ mm/sec$^2$. Hence, in Phantom B, the velocity filter is tested against the acceleration of $\sim0.15$ mm/sec$^2$.

\subsubsection{Velocity Filtering Results}
\begin{figure}[hbt!]
\begingroup
    \centering
    \begin{tabular}{c c c}
\hspace{-0.3cm}{ \includegraphics[width=0.32\linewidth]{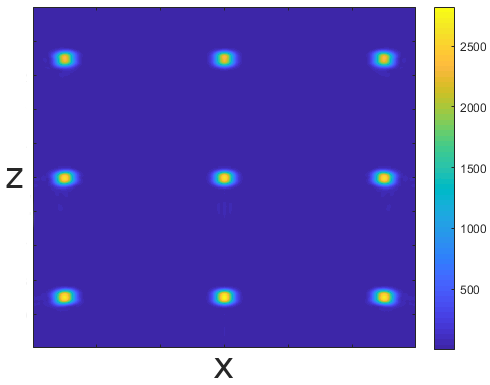}} &\hspace{-0.4cm}{ \includegraphics[width=0.32\linewidth]{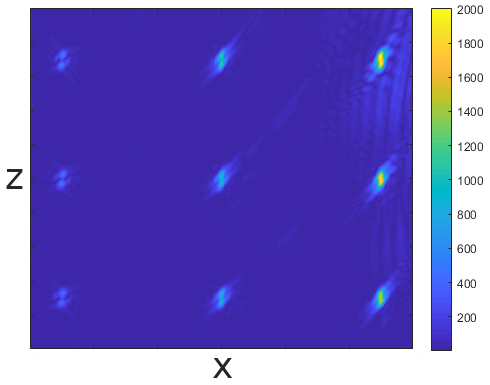}} & \hspace{-0.4cm}{ \includegraphics[width=0.32\linewidth]{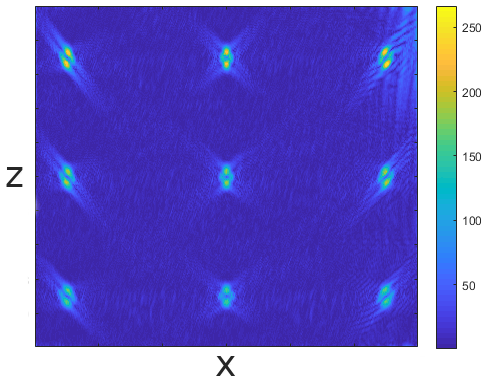}} \\
{\footnotesize \hspace{-0.7cm} (a) $Attn \approx 4.6$} &\hspace{-0.7cm} (b){\footnotesize $Attn \approx 13$}&\hspace{-0.5cm} (c){\footnotesize $Attn \approx 100$ }
\end{tabular}
\caption{Phantom A:  Velocity-filtered  frame magnitude 
(Note the different magnitude scales on the color bars.) 
(a) $\sigma_t = 0.5$ sec, $\vc{\Delta v} = 1 \angle{0^{\circ}}$ mm/sec 
(b) $\sigma_t = 0.1$ sec, $\vc{\Delta v} = 1 \angle{-45^{\circ}}$ mm/sec (c) $\sigma_t = 0.1$ sec, $\vc{\Delta v} = 1 \angle{-90^{\circ}}$ mm/sec.   
}
\label{PhantomAPre}
\endgroup
\end{figure}

 With data from Phantom A as input, Fig.~\ref{PhantomAPre} displays the magnitude of the output of the velocity filter at a particular time instant, for two values of $\sigma_t$ and for three selected directions $\vc{v_f}$,  corresponding to three directions of $\vc{\Delta v}$. The reported attenuation is calculated based on the microbubble located at the center.

The results confirm the predictions of the theoretical analysis.  As predicted by \eqref{eq:q bandpass}, the velocity filter has different attenuation for different $\vc{\Delta v}$ directions.
In Fig.~\ref{PhantomAPre}a, with $\sigma_t = 0.5$sec, we see moderate  attenuation  of $4.6$ for $\vc{\Delta v}$ in the lateral direction. From Figs.~\ref{PhantomAPre}b-c, we see that even with reduced $\sigma_t = 0.1$sec,  as $\vc{\Delta v}$ rotates from diagonal to axial direction, the attenuation increases from  13 to 100 for $\| \vc{\Delta v}\| = 1$ mm/sec. This significantly  improved attenuation with increasing $|{\Delta v}_z|$ was predicted by the exponential decay factor in  \eqref{gammaforpreenv}.

An observation that could not be predicted by the analysis in Section \ref{proposedmethodchapter}, which used a spatially shift-invariant model, is regarding the spatially varying PSF that manifests in practice at larger lateral displacements in the near field. Fig.~\ref{PhantomAPre} shows that the velocity filter has a spatially varying response: microbubbles located on the right side, at the center, and on the left side of the transducer do not show the same response to the velocity filter. This is expected because microbubbles having different lateral positions have PSFs that are relatively rotated around their centers. Thus, the coordinate system in which $\vc{\Delta v}$ is defined is similarly rotated, resulting in different attenuation factors between left, center and right microbubbles that have the same $\vc{\Delta v}$ in absolute coordinates. 

\begin{figure}[hbt!]
\begingroup
    \centering
    \begin{tabular}{c c}
\hspace{-0.4cm}{ \includegraphics[width=0.5\linewidth]{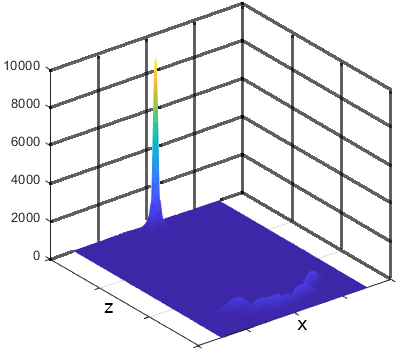}} & \hspace{-0.4cm}{\includegraphics[width=0.5\linewidth]{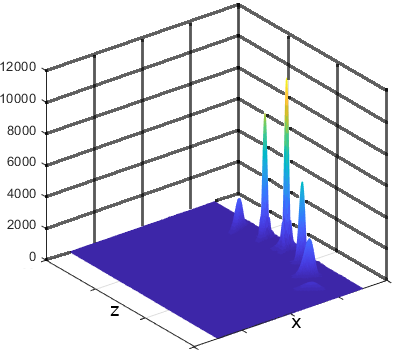}}\\
{\footnotesize \hspace{-0.5cm} $\vc{v_f} = 1 \angle{180^{\circ}}\frac{\mathrm{mm}}{\mathrm{sec}}$}&{\footnotesize \hspace{-0cm}  $\vc{v_f} = 1 \angle{90^{\circ}}\frac{\mathrm{mm}}{\mathrm{sec}}$ }
\end{tabular}
\caption{Phantom B: Velocity-filtered frame magnitude for $\sigma_t = 0.5 sec$.
}
\label{PhantomCPre}
\endgroup
\end{figure}

Fig.~\ref{PhantomCPre} shows the corresponding results for Phantom B.
We observe that the responses of microbubbles moving at $\vc{v_f}$ are well preserved, and bubble responses with particular $\vc{\Delta v}$ are distorted in the $\vc{\Delta v}$ direction and attenuated with respect to $\| \vc{\Delta v}\|$. In the left figure, the suppression of neighboring bubbles is stronger than the ones in the right figure since the axial $\Delta v_z$ component is larger. 

Importantly, we observe that for $\sigma_t =0.5$sec the velocity filter's effect on circular motion is the same as predicted by the constant velocity model. In other words, the velocity filter tolerates well the acceleration of $\sim 0.15$mm/sec$^2$, and therefore we conclude that for reasonable acceleration the velocity filter is as effective as for constant velocity.

\subsubsection{Transverse Oscillation} As discussed before, the velocity filter has weaker attenuation for $\vc{\Delta v}$ with a zero axial direction component. This limitation is overcome by introducing TO using the filter in \eqref{TOfilterspec}. In the numerical experiments, TO is used with parameters $k_{0x} = 2 \pi/ (\lambda /2)$  and $\sigma_x = \lambda$.

\begin{figure}[hbt!]
\begingroup
    \centering
    \begin{tabular}{c c c}
\hspace{-0.3cm}{ \includegraphics[width=0.32\linewidth]{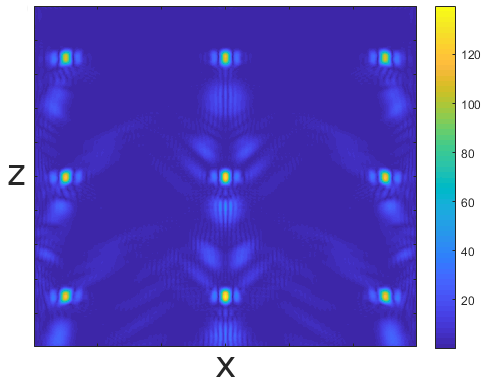}}& \hspace{-0.4
cm}{ \includegraphics[width=0.32\linewidth]{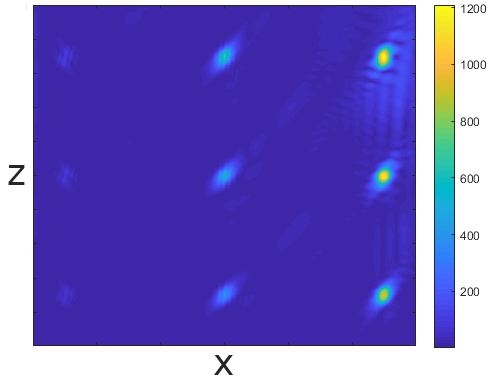}} & \hspace{-0.4cm}{ \includegraphics[width=0.32\linewidth]{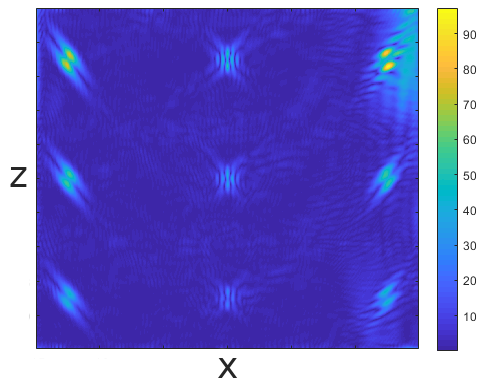}} \\{\footnotesize
\hspace{-0.7cm}  (a)$ Attn \approx 43 $ }&{\footnotesize \hspace{-0.7cm} (b)$Attn \approx 11$}&{\footnotesize \hspace{-0.5cm} (c)$Attn \approx 100$} 
\end{tabular}
\caption{Phantom A: Velocity filtering with TO.  
(a) $\sigma_t = 0.5$ sec, $\vc{\Delta v} = 1 \angle{0^{\circ}}$ mm/sec (b) $\sigma_t = 0.1$ sec, $\vc{\Delta v} = 1 \angle{-45^{\circ}}$ mm/sec (c) $\sigma_t = 0.1$ sec, $\vc{\Delta v} = 1 \angle{-90^{\circ}}$ mm/sec.
}
\label{PhantomATO}
\endgroup
\end{figure}

Fig.~\ref{PhantomATO} displays snapshots of the magnitude of the output of the velocity filter for various $\sigma_t$ and $\vc{\Delta v}$ directions. As in Fig.~\ref{PhantomAPre}, the attenuation reported is for the center bubble.

The results confirm the predictions of the theoretical analysis: (i) TO introduces significant additional attenuation beyond that offered by  velocity filtering alone for lateral $\vc{\Delta v}$. For instance when $\sigma_t = 0.5$ sec and $\vc{\Delta v} = 1 \angle{0^{\circ}}$mm/sec, the attenuation increases from 4.6 in Fig.~\ref{PhantomAPre}-a without TO, to 43 in Fig.~\ref{PhantomATO} with TO. (ii) TO reduces the attenuation for other directions such as -45 degrees. (iii) With TO filtering, the peak signal level for all microbubbles in Fig.~\ref{PhantomATO}, including those moving at the selected velocity, decreases (to half its value, for the example shown). Thus, the results lend further support to the recommendation in Sec.~\ref{TOcasetheory} for the use of TO in the processing pipeline: use TO to reconstruct flow close to later (e.g. at angles $<$ $10^\circ$), but leave it out otherwise. 

\subsection{Numerical Experiments for Localization}
Next, we study, using flow phantoms C-F of Fig.~\ref{Phantom Simulations VF}, the effects of the velocity filter on microbubble localization, recovery of the complete vessel structure, and recovery of the velocity map. The spatially varying template PSF for the matched filter in the localization step is obtained using Field II.

\subsubsection{Phantom C}
Phantom C 
simulates two vessels 
crossing at different (unresolved) cross-plane $y$ positions with perpendicular flow velocities at the vessel center of $ \vc{v} = 5 \angle{\pm45^{\circ}}$ 
(See Tables \ref{tab:my_label_spec_VF}  and \ref{tab:phantomdspecs} for detailed simulation specifications.)

\begin{table}[htb!]
    \centering
    \caption{ \footnotesize simulation specifics for phantom c}
 {\footnotesize
\begin{tabular}{ |p{3.4cm}||p{4.3cm}|  }
\hline
 Simulation Parameters & Values/Expressions \\
 \hline
Simulation Duration &5 seconds\\
Blood Flow Model & Laminar Flow with Parabolic Velocity Profile with peak speed of 5mm/sec\\
Vessel Diameter & $\lambda$\\
Bubble Concentration $C_{MB}$ & 2500 and 400 MBs/mm$^3$\\
Flow Directions $\theta$ & $\pm45$
degrees\\
$\sigma_t$ & 0.5 seconds\\
Velocity Filtering Setting & Pre-Envelope Implementation\\
 \hline
\end{tabular}
}
\label{tab:phantomdspecs}
\end{table}

\begin{figure}[hbt!]
\begingroup
    \centering
    \begin{tabular}{c c c}
\hspace{-0.3cm}{ \includegraphics[width=0.31\linewidth]{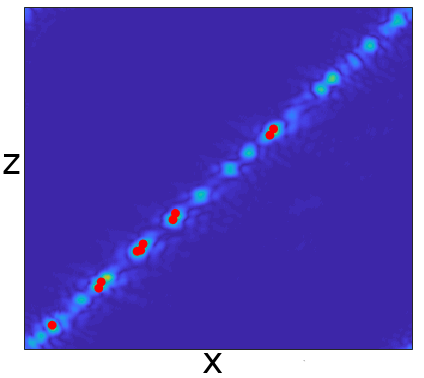}} & \hspace{-0.3cm}{ \includegraphics[width=0.31\linewidth]{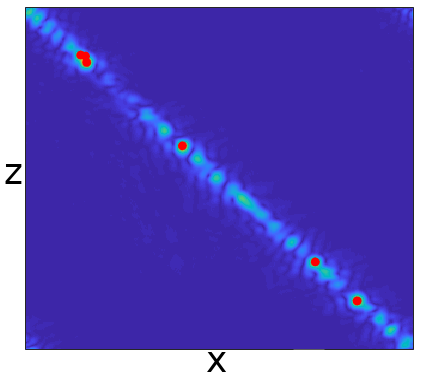}} &\hspace{-0.3cm}{\includegraphics[width=0.36\linewidth]{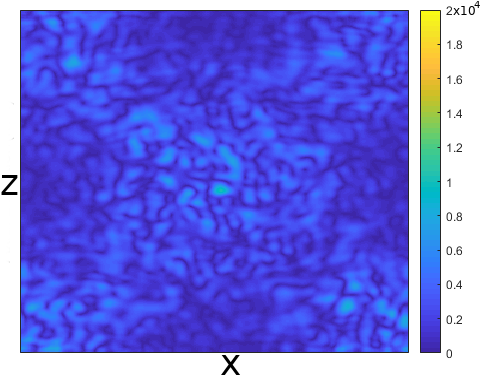}}  \\
\end{tabular}
\caption{Localization results for Phantom C after velocity filter where red dots represent the localized microbubble centers. Each column has different $\vc{v_f}$; $5 \angle{-45^{\circ}}$, $5 \angle{45^{\circ}}$ and  $5 \angle{0^{\circ}}$ mm/sec respectively from left to right column. $C_{MB}$ is 2500MBs/mm$^3$. The colorbars represent the magnitude of the velocity filter output at a particular time instant.}
\label{PhantomDvisual1}
\endgroup
\end{figure}

Fig.~\ref{PhantomDvisual1}, shows velocity filter outputs for three different velocity filter settings $\vc{v_f}$, with  the  corresponding localization results superimposed (in red). It can be seen that the velocity filter successfully selects the microbubbles moving at $\vc{v}=\vc{v_f}$. In the absence of microbubbles moving at the selected velocity $\vc{v_f}$,  the velocity filter attenuates and distorts all the microbubbles. Subsequently, they are easily eliminated by the matched filtering and thresholding step,
resulting in no detected microbubbles.
 
Next, we examine, again using Phantom C, the recovery of vessel structure.

\begin{figure}[hbt!]
\begingroup
    \centering
    \begin{tabular}{c c c}
\hspace{-0.3cm}{ \includegraphics[width=0.31\linewidth]{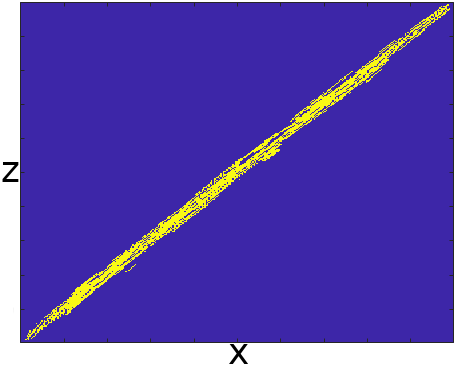}} & \hspace{-0.4cm}{ \includegraphics[width=0.31\linewidth]{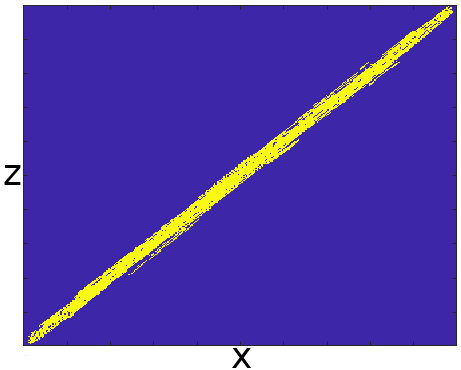}} & \hspace{-0.4cm}{ \includegraphics[width=0.31\linewidth]{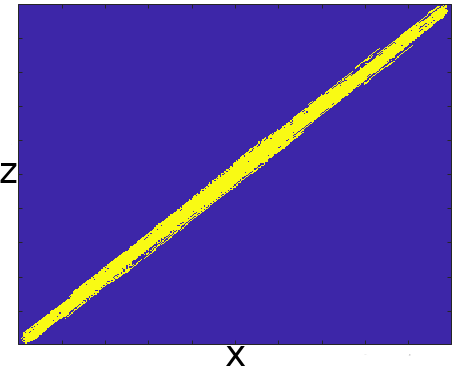}}\\
{\footnotesize \hspace{-0.3cm} $time =0.5$ seconds}& {\footnotesize \hspace{-0.4cm} $time =1$ seconds} & {\footnotesize \hspace{-0.4cm} $time =2$ seconds} \\

\end{tabular}
\caption{Reconstruction of the structure of \emph{vessel 1} in Phantom C with high microbubble density $C_{MB}=2500$ MBs/mm$^3$ using velocity filtering, at different time instants.}
\label{PhantomDvessel1struct}
\endgroup
\end{figure}

Fig.~\ref{PhantomDvessel1struct} illustrates the accumulated microbubble centers at different time instants for high $C_{MB}$, using the velocity filter (VF). It can be seen that the reconstructed vessel support is improving with increasing acquisition time, and at around 2 seconds the reconstructed vessel support become mostly populated by recovered centers. Each of the two overlapping vessel structures in Phantom C can be reconstructed individually with no interference, thanks to the velocity selectivity of the VF. 

For a quantitative analysis of the recovery of vessel structure in Phantom C, we turn to the  $IoU$ metric \eqref{IoUdefinition}. To directly determine the ability of the velocity filter to circumvent the tradeoff between resolution and acquisition (imaging) time, we study the effects of acquisition time and microbubble concentration on $IoU$, with and without velocity filtering.

\begin{figure}[hbt!]
\begingroup
    \centering
    \begin{tabular}{c c}
\hspace{-0.4cm}{ \includegraphics[width=0.51\linewidth]{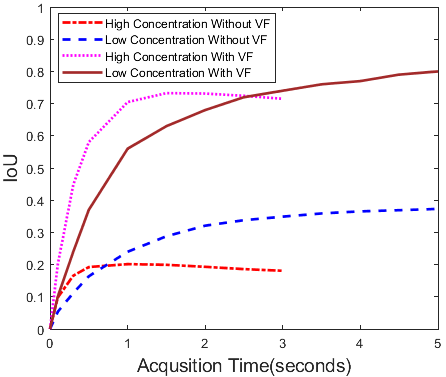}} & 
\hspace{-0.6cm}{ \includegraphics[width=0.51\linewidth]{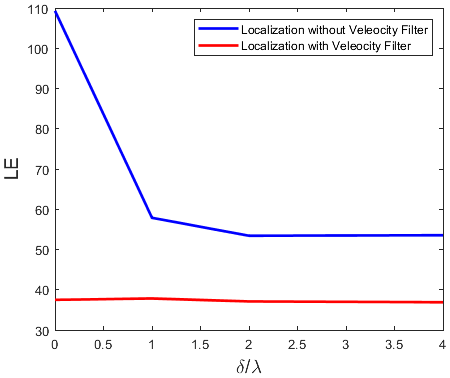}} \\
{\footnotesize \hspace{-1cm}{(a)}}&{\footnotesize \hspace{-1cm}{(b)}}
\end{tabular}
\caption{a) IoU versus acquisition time using high  and low microbubble concentration (2500 MBs/mm$^3$ and 400 MBs/mm$^3$) in Phantom C, with and without velocity filter (VF)  b) $LE$ vs $\delta$.
}
\label{IoUvsAc}
\endgroup
\end{figure}

The $IoU$ vs. acquisition time is plotted in Fig.~\ref{IoUvsAc}-a for two cases: localization with (i) and without (ii) velocity filter (VF), each case at two microbubble concentration levels: high ($C_{MB} = 2500$ MBs/mm$^3$) and low ($C_{MB} = 400$ MBs/mm$^3$). 

Fig.~\ref{IoUvsAc}a demonstrates the superior vessel support recovery with the velocity filter.  For example, at high $C_{MB}$, with VF the $IoU$ reaches $\sim 0.7$, whereas without VF the $IoU$ saturates at $\sim 0.2$. The reason for this improvement is that without VF localized microbubble centers are located mostly outside the true vessel boundaries. 

Importantly, Fig.~\ref{IoUvsAc}a demonstrates 
that VF provides a significant reduction in acquisition time, with superior resolution. For high $C_{MB}$, the $IoU$ reaches $\sim 0.7$ at 1.5 sec with VF. In contrast, the same level of $IoU$ cannot be achieved without VF even by decreasing $C_{MB}$  to 400 MBs/mm$^3$: without VF, the $IoU$  only reaches $\sim 0.4$ at 5 sec.


\subsubsection{Phantom D}
The phantom consists of two parallel vessels with flow in opposite directions separated by a small gap $\delta$.  The simulation and localization settings are the same as for Phantom C, except $C_{MB}=2500$MBs/mm$^3$. The goal of this study is to evaluate the ability of the velocity filtering to help resolve small tightly packed microvessels, using velocity differences. To this end, we evaluate the localization error metric $LE$ \eqref{LEmetric}  vs. the inter-vessel gap $\delta$, with lower $LE$  meaning better localization accuracy. The $LE$ parameters 
are chosen as $\sigma_\|=0.3 \lambda$ and $\sigma_\perp=0.15 \lambda$. For each $\delta$, $LE$ is calculated at a fixed time instant with and without VF. 

The $LE$ plots in Figure \ref{IoUvsAc}-b  show 
that as the gap $\delta$ between the two vessels diminishes to below $\lambda$,
without VF the $LE$ increases sharply, suggesting that the vessels would no longer be resolved. In contrast, with VF, the localization accuracy is maintained even when the two vessels start to overlap.           

\subsubsection{Phantom E}
The goal of this study is to evaluate the recovery of flow velocity information and the flow velocity map using the velocity filter. To this end, Phantom E simulates a single 3D vessel, and the entire processing chain with VF described in the Introduction is used to recover the parabolic velocity profile. The simulation specifics are the same as for Phantom C, including the localization settings with 4 sec. accumulation time, except the following: (i) the vessel diameter  $d=4 \lambda$; (ii) the flow has velocity $ \vc{v} = 5 \angle{-45^{\circ}}$ at the vessel center; and
 (iii) $C_{MB}=225$ MBs/mm$^3$. 

\begin{figure}[hbt!]
\begingroup
    \centering
    \begin{tabular}{c c}
{ \includegraphics[width=0.42\linewidth]{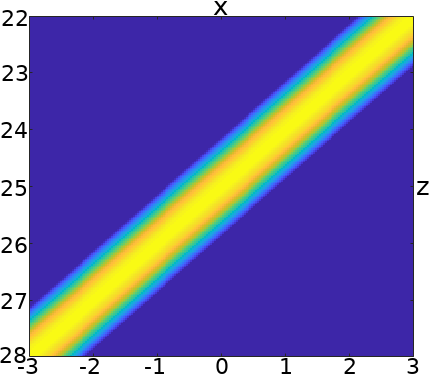}}&
{ \includegraphics[width=0.45\linewidth]{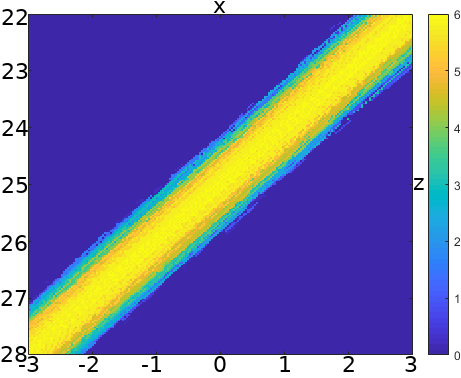}}\\
{\footnotesize 
(a) Ground Truth } &{\footnotesize 
(b) Recovered Speed Map} \\
\end{tabular}
\caption{True and estimated flow velocity maps for
Phantom E. $FVE = 0.56$ mm/sec per pixel. The colorbar represents speed in mm/sec. The numbers on the axes are in mm.}
\label{phontomesimres}
\endgroup
\end{figure}

The simulation, true to the physical model with a 1D ultrasound array and 3D flow discussed earlier, projects all the scatters along the elevation direction $y$ onto the beamformed image plane, and hence, multiple speed values might be recovered for one image pixel.  The recovered map in Fig.~\ref{phontomesimres} is obtained by assigning the maximum speed of microbubbles detected in the pixel. Hence, the speed map of the vessel's center plane is recovered, and therefore a parabolic velocity profile is observed, as expected. 

For a quantitative assessment of the accuracy of the recovery of the flow velocity map, we computed the flow velocity error metric \eqref{FVEdef}. The resulting value of $FVE = 0.56$ mm/sec indicates good accuracy, especially when combined with the super-resolved velocity distribution. Restricting the assessment to 
the fastest 5\% of the bubbles yields 
a flow velocity error $FVE = 0.19$ mm/sec, which is $\approx 3\%$ of the speed at the center of the vessel.

\subsubsection{Phantom F}
This phantom simulates circular flow with a radius of 7.5mm.  The simulation specifics are the same as in Phantom C including the localization, except that (i) the vessel diameter is $3\lambda$;
(ii) the flow speed at the vessel center is $2$mm/sec; and  $C_{MB}= 100$MBs/mm$^3$. With these parameters, the maximum centripetal acceleration experienced by microbubbles is $\sim 0.5$ mm/sec$^2$. Because the circular flow violates the constant velocity assumption, it provides a critical test for overall processing pipeline including localization. 

\begin{figure}[hbt!]
\begingroup
    \centering
    \begin{tabular}{c c c}
\hspace{-0.3cm}{ \includegraphics[width=0.30\linewidth]{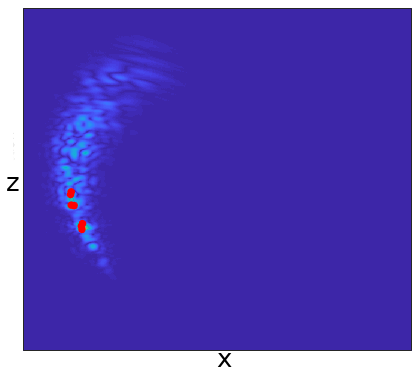}} & \hspace{-0.5cm}{ \includegraphics[width=0.30\linewidth]{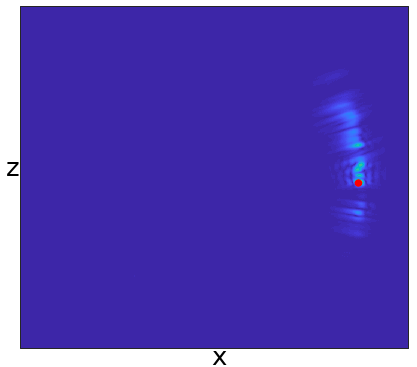}} &\hspace{-0.5cm}{ \includegraphics[width=0.34\linewidth]{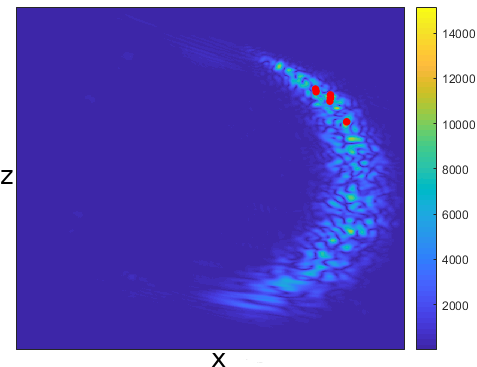}} \\
{\footnotesize \hspace{-0.3cm}  $\vc{v_f} = 2 \angle{60^{\circ}}\frac{\mathrm{mm}}{\mathrm{sec}}$}& {\footnotesize\hspace{-0.5cm} $\vc{v_f} = 2\angle{-90^{\circ}}\frac{\mathrm{mm}}{\mathrm{sec}}$}&{\footnotesize \hspace{-0.5cm} $\vc{v_f} = 2 \angle{-150^{\circ}}\frac{\mathrm{mm}}{\mathrm{sec}}$}\\
 \hspace{-0.3cm}{ \includegraphics[width=0.30\linewidth]{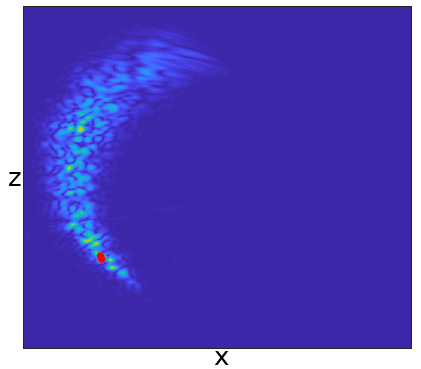}} & \hspace{-0.5cm}{\includegraphics[width=0.33\linewidth]{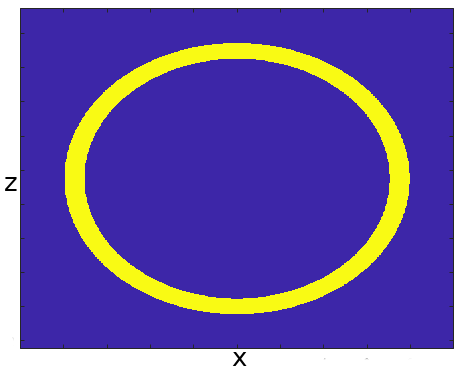}}& \hspace{-0.5cm}{ \includegraphics[width=0.33\linewidth]{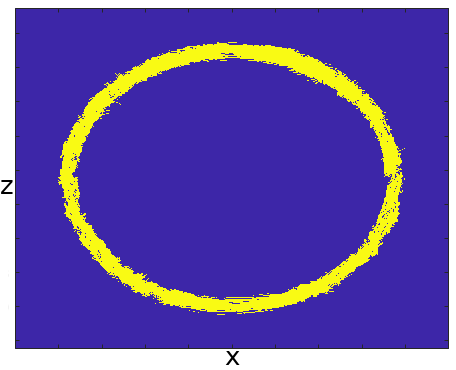}} \\ {\footnotesize \hspace{-0.3cm} $\vc{v_f} = 2\angle{45^{\circ}}\frac{\mathrm{mm}}{\mathrm{sec}}$}&{\footnotesize\hspace{-0.5cm} Ground Truth} &{\footnotesize \hspace{-0.5cm} $time =3$ seconds}
 \\
\end{tabular}
\caption{Phantom F -- circular flow recovery with the velocity filter. Top row and left bottom: VF outputs (magnitude represented by the color bar) with superimposed localizations (in red). Bottom middle and right:  ground truth vessel support and accumulated microbubble centers.}
\label{PhantomGlocal}
\endgroup
\end{figure}

 The velocity filter outputs for several $\vc{v_f}$ are displayed in in Fig.~\ref{PhantomGlocal}, along with localized microbubble centers superimposed as red dots. It can be seen that the microbubble centers are successfully localized for the circular motion in spite of the substantial deviation from constant velocity motion.  The reconstructed vessel support at 3 seconds of accumulation displayed in Fig.~\ref{PhantomGlocal} is in quite good visual agreement with the ground truth vessel support. This agreement is quantified by computing the $IoU$ metric,  $IoU=0.82$.

\vspace*{-0.1cm}
\section{Conclusions}
\label{chapter6}
ULM offers a cost-effective and non-invasive modality for microvascular imaging, however its clinical translation has been challenging because of its inherent trade-off between acquisition time and resolution. The velocity filtering introduced in this paper is a simple yet powerful method to incorporate microbubble dynamics into the localization procedure itself, and thus
circumvent this tradeoff, providing superresolution with a short acquisition time. As an important side benefit, thanks to its narrow-band spatio-temporal filtering prior to detection and localization (analogous to so-called pre-detection integration, in radar), the proposed velocity filter has the potential to extend the penetration depth of ULM in tissue by 4 cm or more.
Along with velocity selective localization and precise microvascular imaging, the filtering operation simultaneously provides blood velocity mapping as well, eliminating the need for complex subsequent micro-bubble tracking and speed estimation steps.
Thanks to the simplicity and low computational cost of the method, 
it is amenable to real-time implementation. We therefore  believe that with these multitude advantages, using velocity filtering as a preprocessing step before localization will pave the way to its clinical translation.  


In future work, the technique will be tested on real data. It will also be interesting to consider the generalizations to three-dimensional (3D) imaging, and a 6D phase space.


\section*{Acknowledgment}
The authors thank Prof. Pengfei Song for helpful discussions.
\bibliographystyle{IEEE_ECE}
\bibliography{thesisrefs}

\interlinepenalty10000
\begin{thebibliography}{10}
\providecommand{\url}[1]{#1}
\csname url@samestyle\endcsname
\providecommand{\newblock}{\relax}
\providecommand{\bibinfo}[2]{#2}
\providecommand{\BIBentrySTDinterwordspacing}{\spaceskip=0pt\relax}
\providecommand{\BIBentryALTinterwordstretchfactor}{4}
\providecommand{\BIBentryALTinterwordspacing}{\spaceskip=\fontdimen2\font plus
\BIBentryALTinterwordstretchfactor\fontdimen3\font minus
  \fontdimen4\font\relax}
\providecommand{\BIBforeignlanguage}[2]{{%
\expandafter\ifx\csname l@#1\endcsname\relax
\typeout{** WARNING: IEEEtran.bst: No hyphenation pattern has been}%
\typeout{** loaded for the language `#1'. Using the pattern for}%
\typeout{** the default language instead.}%
\else
\language=\csname l@#1\endcsname
\fi
#2}}
\providecommand{\BIBdecl}{\relax}
\BIBdecl

\bibitem{book}
T.~L. Szabo, \emph{Diagnostic Ultrasound Imaging: Inside Out}.\hskip 1em plus
  0.5em minus 0.4em\relax Academic Press, 2004.

\bibitem{padhani2005angiogenesis}
A.~R. Padhani, C.~J. Harvey, and D.~O. Cosgrove, ``Angiogenesis imaging in the
  management of prostate cancer,'' \emph{Nature Clinical Practice Urology},
  vol.~2, no.~12, pp. 596--607, 2005.

\bibitem{jia2016three}
W.-R. Jia~et al., ``Three-dimensional contrast-enhanced ultrasound in response
  assessment for breast cancer: A comparison with dynamic contrast-enhanced
  magnetic resonance imaging and pathology,'' \emph{Scientific Reports},
  vol.~6, p. 33832, 2016.

\bibitem{deffieux2018functional}
T.~Deffieux, C.~Demene, M.~Pernot, and M.~Tanter, ``Functional ultrasound
  neuroimaging: A review of the preclinical and clinical state of the art,''
  \emph{Current Opinion in Neurobiology}, vol.~50, pp. 128--135, 2018.

\bibitem{errico2015ultrafast}
C.~Errico~et al., ``Ultrafast ultrasound localization microscopy for deep
  super-resolution vascular imaging,'' \emph{Nature}, vol. 527, no. 7579, pp.
  499--502, 2015.

\bibitem{viessmann2013acoustic}
O.~M. Viessmann, R.~J. Eckersley, K.~Christensen-Jeffries, M.~X. Tang, and
  C.~Dunsby, ``Acoustic super-resolution with ultrasound and microbubbles,''
  \emph{Phys. Med. Biol.}, vol.~58, no.~18, p. 6447, 2013.

\bibitem{christensen2014vivo}
K.~Christensen-Jeffries, R.~J. Browning, M.~X. Tang, C.~Dunsby, and R.~J.
  Eckersley, ``In vivo acoustic super-resolution and super-resolved velocity
  mapping using microbubbles,'' \emph{IEEE Trans. Med. Im.}, vol.~34, no.~2,
  pp. 433--440, 2014.

\bibitem{betzig2006imaging}
E.~Betzig~et al., ``Imaging intracellular fluorescent proteins at nanometer
  resolution,'' \emph{Science}, vol. 313, no. 5793, pp. 1642--1645, 2006.

\bibitem{rust2006sub}
M.~J. Rust, M.~Bates, and X.~Zhuang, ``Sub-diffraction-limit imaging by
  stochastic optical reconstruction microscopy ({STORM}),'' \emph{Nature
  Methods}, vol.~3, no.~10, pp. 793--796, 2006.

\bibitem{song2017improved}
P.~Song~et al., ``Improved super-resolution ultrasound microvessel imaging with
  spatiotemporal nonlocal means filtering and bipartite graph-based microbubble
  tracking,'' \emph{IEEE Trans. Ultras. Ferroel. Freq. Cont.}, vol.~65, no.~2,
  pp. 149--167, 2017.

\bibitem{bar2018sushi}
A.~Bar-Zion, O.~Solomon, C.~Tremblay-Darveau, D.~Adam, and Y.~C. Eldar,
  ``Sushi: Sparsity-based ultrasound super-resolution hemodynamic imaging,''
  \emph{IEEE Trans. Ultras. Ferroel. Freq. Cont.}, vol.~65, no.~12, pp.
  2365--2380, 2018.

\bibitem{solomon2019exploiting}
O.~Solomon, R.~J. Van~Sloun, H.~Wijkstra, M.~Mischi, and Y.~C. Eldar,
  ``Exploiting flow dynamics for superresolution in contrast-enhanced
  ultrasound,'' \emph{IEEE Trans. Ultras. Ferroel. Freq. Cont.}, vol.~66,
  no.~10, pp. 1573--1586, 2019.

\bibitem{tang2020kalman}
S.~Tang~et al., ``Kalman filter--based microbubble tracking for robust
  super-resolution ultrasound microvessel imaging,'' \emph{IEEE Trans. Ultras.
  Ferroel. Freq. Cont.}, 2020.

\bibitem{van2018super}
R.~J. Van Sloun~et al., ``Super-resolution ultrasound localization microscopy
  through deep learning,'' \emph{arXiv:1804.07661}, 2018.

\bibitem{huang2020short}
C.~Huang~et al., ``Short acquisition time super-resolution ultrasound
  microvessel imaging via microbubble separation,'' \emph{Scientific Reports},
  vol.~10, no.~1, pp. 1--13, 2020.

\bibitem{yang2017moving}
L.~Yang, L.~Yang, and K.~Ho, ``Moving target localization in multistatic sonar
  using time delays, doppler shifts and arrival angles,'' in \emph{ICASSP
  2017}, pp. 3399--3403.

\bibitem{montaldo2009coherent}
G.~Montaldo, M.~Tanter, J.~Bercoff, N.~Benech, and M.~Fink, ``Coherent
  plane-wave compounding for very high frame rate ultrasonography and transient
  elastography,'' \emph{IEEE Trans. Ultras. Ferroel. Freq. Cont.}, vol.~56,
  no.~3, pp. 489--506, 2009.

\bibitem{ivanov1981blood}
K.~Ivanov, M.~Kalinina, and Y.~I. Levkovich, ``Blood flow velocity in
  capillaries of brain and muscles and its physiological significance,''
  \emph{Microvasc. Res.}, vol.~22, no.~2, pp. 143--155, 1981.

\bibitem{hoskins2000haemodynamics}
P.~R. Hoskins, W.~N. McDicken, and P.~L. Allan, ``Haemodynamics and blood
  flow,'' \emph{Clinical Doppler Ultrasound}, pp. 27--38, 2000.

\bibitem{desailly2015resolution}
Y.~Desailly, J.~Pierre, O.~Couture, and M.~Tanter, ``Resolution limits of
  ultrafast ultrasound localization microscopy,'' \emph{Phys. Med. Biol.},
  vol.~60, no.~22, p. 8723, 2015.

\bibitem{chen2020algorithmic}
S.~Chen and A.~Moitra, ``Algorithmic foundations for the diffraction limit,''
  \emph{arXiv preprint arXiv:2004.07659}, 2020.

\bibitem{hingot2019microvascular}
V.~Hingot, C.~Errico, B.~Heiles, L.~Rahal, M.~Tanter, and O.~Couture,
  ``Microvascular flow dictates the compromise between spatial resolution and
  acquisition time in ultrasound localization microscopy,'' \emph{Scientific
  Reports}, vol.~9, no.~1, pp. 1--10, 2019.

\bibitem{varray2013alternative}
F.~Varray and H.~Liebgott, ``An alternative method to classical beamforming for
  transverse oscillation images: Application to elastography,'' in \emph{ISBI
  2013}, pp. 716--719.

\bibitem{demene2015spatiotemporal}
C.~Demen{\'e}~et al., ``Spatiotemporal clutter filtering of ultrafast
  ultrasound data highly increases {D}oppler and fultrasound sensitivity,''
  \emph{IEEE Trans. Med. Im.}, vol.~34, no.~11, pp. 2271--2285, 2015.

\bibitem{michailovich2006despeckling}
O.~V. Michailovich and A.~Tannenbaum, ``Despeckling of medical ultrasound
  images,'' \emph{IEEE Trans. Ultras. Ferroel. Freq. Cont.}, vol.~53, no.~1,
  pp. 64--78, 2006.

\bibitem{jensen1996field}
J.~A. Jensen, ``Field: A program for simulating ultrasound systems,'' in
  \emph{10th Nordicbaltic Conf. Biomed. Imag.}, vol.~4, 1996, pp. 351--353.

\end{thebibliography}
\def\x{{\mathbf x}}
\def\L{{\cal L}}
\renewcommand{\Re}{\operatorname{Re}}
\renewcommand{\Im}{\operatorname{Im}}
\newcommand{\Rtwo}{\mathbb{R}^2}
\setcounter{equation}{0}
\newcounter{Cequ}
\newcounter{Caux}
\setcounter{Cequ}{0}
\setcounter{Caux}{0}
\newenvironment{CAlign}
   {\setcounter{Caux}{\theequation}
     \setcounter{equation}{\theCequ}%
     \renewcommand\theequation{A\arabic{equation}}
     \align}
   {\endalign\setcounter{Cequ}{\value{equation}}\setcounter{equation}{\theCaux}}

\section*{Appendix}
\subsection{Velocity Filter Derivation}
\subsubsection{Microbubble Signal $B(\vc{k},\Omega)$}
\label{AppendixA Microbubble}
 Taking the Fourier transform $\F_{\vc{r} \rightarrow \vc{k}} \{b(\vc{r},t)\}$ of the spatio-temporal ULM data \eqref{bubblesignal},
\begin{CAlign}
\label{apendixbublekt}
    B(\vc{k},t) = G(\vc{k}) \exp{(-\vc{k} \cdot (\vc{r_0}+\vc{v}t))}
\end{CAlign}
Then another Fourier transform $\F_{t \rightarrow \Omega} \{B(\vc{k},t)\}$  results in 
\begin{CAlign}
    B(\vc{k},\Omega) = G(\vc{k}) \exp{(-i \vc{k} \cdot \vc{r_0})} \delta(\Omega +\vc{k} \cdot \vc{v} )
\end{CAlign}
\subsubsection{Effective Impulse Response $\hat{h}(\vc{r}, \vc{\Delta v})$}
\label{appendixintegration}
The impulse response of the effective velocity filter $\hat{H}(\vc{k} , \vc{\Delta v }) \triangleq W(\vc{k} \cdot \vc{\Delta v }) $ corresponding to \eqref{eq:PhikO}
is calculated by change of variables in
\[\hat{h}(\vc{r}, \vc{\Delta v}) = \left(\frac{1}{2 \pi}\right)^2 \int_{\Rtwo} 
 W[\vc{k} \cdot  \vc{\Delta v}] \exp(i \vc{k} \cdot \vc{r}) d\vc{k} \]
Let $\vc{u_1} \defsign \hat{\vc{\Delta v}} = \vc{\Delta v}/|| \vc{\Delta v} ||$, $\vc{u_2} = \hat{\vc{\Delta v}}^\perp  $, $\vc{u_1} \cdot \vc{u_2} = 0$, be orthonormal vectors in the directions of $\vc{\Delta v}$ and perpendicular to it, respectively. Then $\vc{U} = \begin{bmatrix} 
 \vc{u_1}, \vc{u_2} 
 \end{bmatrix} \in \R ^{2 \times 2} $ is an orthogonal matrix, $U^T U = U U^T = I$. Making the change of variables $ \vc{k} = \vc{U} \vc{q}$ 
 yields
\[\vc{k} \cdot \vc{\Delta v} = \vc{q}^T \vc{U}^T \vc{\Delta v} = \begin{bmatrix} q_1, q_2
\end{bmatrix} \begin{bmatrix} \hat{\vc{\Delta v}} \cdot \vc{\Delta v} \\ \hat{\vc{\Delta v}}^\perp \cdot \vc{\Delta v}  \end{bmatrix}  = q_1 || \vc{\Delta v}||\]
\[\vc{k} \cdot \vc{r} = \vc{q}^T \vc{U}^T \vc{r} =  q_1 \hat{ \vc{\Delta v}} \cdot \vc{r} + q_2 \hat{ \vc{\Delta v}}^\perp \cdot \vc{r}\]
\[ d\vc{k} = dk_1dk_2 = dq_1dq_2\]
Hence, the integral becomes
\[
\int_{\Rtwo} 
 W(q_1 || \vc{\Delta v}||) e^{i q_1 \hat{ \vc{\Delta v}} \cdot \vc{r} + q_2 \hat{ \vc{\Delta v}}^\perp \cdot \vc{r} } d\vc{q} 
 \]
Defining $ \alpha \defsign q_1 || \vc{\Delta v}|| $ 
this reduces to
\begin{CAlign}
     &= \left(\frac{1}{|| \vc{\Delta v}|| }\right) \int\limits_{- \infty}^{\infty}
 W(\alpha ) e^{i \alpha \frac{\hat{ \vc{\Delta v}} \cdot \vc{r}}{|| \vc{\Delta v}|| }  }d\alpha  \int\limits_{- \infty}^{\infty} e^{i q_2 \hat{ \vc{\Delta v}}^\perp \cdot \vc{r} }dq_2 \nonumber \\
 &=[(2 \pi)^2/|| \vc{\Delta v}||] w\left(\hat{ \vc{\Delta v}} \cdot \vc{r}/|| \vc{\Delta v}||  \right) \delta(\hat{ \vc{\Delta v}}^\perp \cdot \vc{r} ) \nonumber 
\end{CAlign}
 It follows that 
 \begin{CAlign}
 \label{eq:h-hat}
     \hat{h}(\vc{r}, \vc{\Delta v}) = \begin{cases} \frac{1}{|| \vc{\Delta v}||} w\left(\frac{\hat{ \vc{\Delta v}} \cdot \vc{r}}{|| \vc{\Delta v}|| } \right) \delta(\hat{ \vc{\Delta v}}^\perp \cdot \vc{r} ) & {\scriptstyle|| \vc{\Delta v}|| \not = 0}\\ \delta(\vc{r})& \vc{\Delta v} =0\end{cases}
 \end{CAlign} 
\subsubsection{Pre-Envelope Implementation}
\label{pre-envelopeappendix}
The motion-free filtered microbubble signal is
\begin{CAlign}
q(\vc{r};\vc{\Delta v}) &= g(\vc{r}) \ast_{r} \hat{h}(\vc{r},\vc{\Delta v})  \nonumber\\
&=\begin{cases} \int\limits_{\mathbb{R}^2} \frac{g(\vc{r} - \vc{p})}{|| \vc{\Delta v}||} w(\frac{\hat{ \vc{\Delta v}} \cdot \vc{p} }{|| \vc{\Delta v}|| } ) \delta(\hat{ \vc{\Delta v}}^\perp \cdot \vc{p} )d\vc{p} &{|| \vc{\Delta v}|| \not = 0}\\ g(\vc{r})& \vc{\Delta v} =0 \end{cases} \nonumber
\end{CAlign}
Using the change of variables $\vc{p} = \vc{U} \vc{m}$   with the same orthonormal matrix $\vc{U}$ defined in Appendix A.2 yields 
\begin{CAlign}
\label{eq:q-g-conv-h}
  q(\vc{r};\vc{\Delta v}) = \int\limits_{- \infty}^{\infty}g(\vc{r} - t \vc{\Delta v})  w(t ) dt .
\end{CAlign}
Combining with the pre-envelope PSF \eqref{passbandPSF} and the Gaussian window $w(t)$ the integral becomes
\begin{CAlign}
C
\Re\int\limits_{- \infty}^{\infty}
\exp\left[ 
- \frac{\|\vr - t \delv\|^2}{2 \sigma_r^2}
+ i \frac{2 \pi (z-t \vc{\Delta v}_z)}{ \lambda} -\frac{t^2}{2\sigma_t^2}
\right]  dt \nonumber
\end{CAlign}
where 
\vspace*{-0.2cm}
\[C = \frac{1}{2 \pi \sigma_r^2 \sqrt{2 \pi}\sigma_t } \]
Denoting the  exponent in the integrand by $\psi(t)$
and completing the square we obtain
\begin{CAlign} \label{eq:exponent}
    \psi(t) = \gamma  -\frac{(t-\mu)^2}{2 \sigma_{bw}^2}
\end{CAlign}
where
\vspace*{-0.2cm}
\begin{align*}
\sigma_{bw}^2 & \defsign \left(
\frac{1}{\sigma_t^2} +  \frac{\delv^T \delv}{\sigma_r^2}\
\right)^{-1}
\\
\mu &\defsign \sigma_{bw}^2 \left( \frac{1}{\sigma_r^2}\vr^T \delv - i \frac{ 2\pi \vc{\Delta v}_z}{\lambda} \right) \\
\gamma & \defsign-\frac{\vr^T \vr}{2 \sigma_r^2} + \frac{1}{2} \sigma_{bw}^2
\left( \frac{\vr^T  \delv}{\sigma_r^2}\right)^2 - \frac{1}{2} \sigma_{bw}^2 \left(\frac{ 2\pi \vc{\Delta v}_z}{\lambda}\right)^2 
\\
& \quad + i\frac{ 2\pi }{ \lambda} \left( z- \frac{\sigma_{bw}^2 \vc{\Delta v}_z  \vr^T \delv}{\sigma_r^2}  \right)
\end{align*}
Using \eqref{eq:exponent} 
\[
q(\vc{r},\vc{\Delta v})=  \frac{1}{2 \pi \sigma_r^2  \sqrt{2 \pi}\sigma_t }  \Re \left\{ e^{\gamma} \int_{- \infty}^{\infty}
e^{ -\frac{(t - \mu)^2}{2 \sigma_{bw}^2}} dt \right\} \\
\]
Recognizing the integral  as the mass of an unnormalized Gaussian density
it equals the constant $ \sigma_{bw} \sqrt{2 \pi}$. Hence
\begin{equation*} \label{eq:finalq}
    q(\vc{r},\vc{\Delta v}) = \frac{\sigma_{bw}}{ 2 \pi \sigma_r^2 \sigma_t }  \Re\left\{ \exp( \gamma ) \right\}
\end{equation*}
Next, we determine  the attenuation and distortion relations.
\begin{align*}
    \sigma_{bw}  &= \frac{ \sigma_r \sigma_t}{\sqrt{\sigma_r^2 +  \sigma_t^2 \| \vc{\Delta v }\|^2}} 
    =  \frac{\sigma_{w}}{\sqrt{1+ \kappa^2}}\\
    \kappa & \defsign (\sigma_t/\sigma_r) \| \vc{\Delta v}\|
\end{align*}
\begin{align*}
    \gamma &= - \frac{ \|\vr\|^2}{2\sigma_r^2} \left[ 1- \frac{\kappa^2 \cos^2 (\theta_{\vc{r}, \delv} ) }{1+\kappa^2} \right]  - \frac{2\pi^2}{1+ \kappa^2} \left(\frac{  \sigma_t \vc{\Delta v}_z}{\lambda } 
\right)^2  \nonumber \\ & \quad +i\frac{ 2\pi z }{ \lambda}- i\zeta(\vc{r},\delv)
\end{align*}
where $\theta_{\vc{r}, \delv}$ is the angle between $\vc{r}$ and $\delv$. 
It follows that
\begin{CAlign} \label{finalqforI}
    q(\vc{r},\vc{\Delta v}) =& \Gamma(\vc{\Delta v}) g_e(\eta \vc{r}) \cos \left(\frac{ 2\pi }{ \lambda} (z - \zeta( \vc{r},\delv)) \right)
\end{CAlign}
where
\vspace*{-0.4cm}
\begin{align*}
    \Gamma(\vc{\Delta v}) & \defsign \frac{1}{\sqrt{1 + \kappa^2 }} \exp\left[- \frac{2\pi^2}{1+ \kappa^2} \left(\frac{  \sigma_t \vc{\Delta v}_z}{\lambda }
\right)^2 \right] \\
\eta & \triangleq \sqrt{1-\frac{\kappa^2}{1+\kappa^2}\cos^2(\theta_{\vc{r}, \vc{\Delta v}})} \\
\zeta(\vc{r},\vc{\Delta v}) & \defsign   \frac{\kappa}{(1+\kappa^2)} \frac{\sigma_t \vc{\Delta v}_z} {\sigma_r} \cos (\theta_{\vc{r}, \vc{\Delta v}} ) \|\vc{r}\|
\end{align*}

\subsubsection{Post-Envelope Implementation}
\label{appendpostenvelope}
Because the post-envelope PSF  \eqref{basebandPSF} is missing the cosine  in \eqref{passbandPSF}, 
$q(\vc{r},\vc{\Delta v})$ for this case is simply obtained by setting $i=0$ in the derivation of the pre-envelope case. It follows that  in this case
\begin{CAlign}
\label{motionfreeappendixpost}
q(\vc{r},\vc{\Delta v})    =
   g_e\left(\eta\vc{r}\right)/\sqrt{1 + \kappa^2} .
\end{CAlign}

\subsection{Apparent Microbubble Density}
\label{appendapparentMBdens}
We derive $d_2(v,\rho)$ first. The maximum value of $y$ in the vessel for a given radial distance $\rho$ in the $(x,z)$ plane from the vessel center is given by
\begin{CAlign}
    y_{\max}(\rho) = \sqrt{R^2- \rho^2}
\end{CAlign}
 In the 3D blood flow model, microbubbles along the elevation direction will be projected to the 2D image plane. This can be represented as integration over the elevation coordinate $y$. On the other hand, the same quantity can be obtained by integrating $d_2(v,\rho)$ with respect to $v$. This identity is expressed as
\begin{CAlign}
\label{d2vzintegral}
    \int\limits_{0}^{v_{\max}(\rho)} d_2(v,\rho) dv = \int\limits_{y_{\max}(\rho)}^{-y_{\max}(\rho)} C_{MB}dy
\end{CAlign}
where $v_{\max}(\rho)$ is the maximum velocity at radial position $\rho$ given by \eqref{eq:vmax}. 
Another useful relation is for  the speed $v(\rho,y)$ at position $(\rho, y)$ in the vessel cross section. For the  parabolic velocity profile, we have
\begin{CAlign}
\label{vtoyeq}
    v(\rho, y) = v_{0}\left( 1 - \frac{y^2 + \rho^2 }{R^2}\right) .
\end{CAlign}
This implies 
\vspace*{-0.5cm}
\begin{CAlign}
\label{dvtody}
    dv(\rho, y) = \frac{-2y v_{0}}{R^2} dy
\end{CAlign}
Next we apply a  change of variables  to the right hand side of
\eqref{d2vzintegral}, expressing $y$ in terms of $v=v(\rho, y)$. The limits follow from \eqref{vtoyeq}:  
\begin{equation*}
v(\rho, y) =
    \begin{cases}
     v_{\max}(\rho) & y=0 \\
     0 & y = \pm y_{max}(\rho)
    \end{cases}
\end{equation*}
After change variables, \eqref{d2vzintegral} has integrals with the same limits on both sides, yielding the identity of the integrands
\begin{equation*}
    d_2(v,\rho)dv  = C_{MB} \frac{R^2}{ y v_0} dv
    \end{equation*}
 Solving for $y$ in terms of $v$ from \eqref{vtoyeq}, and using \eqref{eq:vmax} produces the final result
 \begin{CAlign}
  d_2(v,\rho)
    &= \frac{C_{MB} R}{v_{0}} \left[
    \left( 
    1- \frac{v}{v_{\max}(\rho)}
    \right)
    \left( 
    1-\frac{\rho^2}{R^2}
    \right)
    \right]^{-\frac{1}{2}}  \\
    & \quad \times 
    \rect\left(
    \frac{v}{v_{\max}(\rho)}- \frac{1}{2} \right)
     \rect\left(\frac{\rho}{2R}\right) 
     \nonumber
    \end{CAlign}
The $\rect(\cdot)$ functions are included, because the equation is only valid for   ranges of $\rho$ and $v \geq 0$ that correspond to the interior of the vessel. It is readily verified that as expected, integrating $d_2(v,\rho)$ over speed from $0$ to $v_{\max}(\rho)$  produces the same $d_{2}(\rho)$ inside the vessel boundaries as \eqref{densitymodel}.

Next, the apparent bubble density $d_{VF}(v_f,\rho)$ after filtering by the velocity filter tuned to $v_f$   is found by integrating $d_2(v,\rho)$ over the velocity pass-band $R(v_f)$ defined in Section \ref{secapparentmicrobubblechapter4}, yielding
\begin{CAlign}
\frac{d_{VF}(v_f,\rho) }{2 C_{MB} R}    &= \sum_{j=0}^1
\sqrt{\left( 1 - \frac{v_f - (-1)^j\delta v}{v_{0}(\rho)} \right)\left(1 -\left(\frac{\rho}{R}\right)^2 \right)} \nonumber \\
& \times
\rect \left(\frac{v_f - (-1)^j\delta v }{v_0(\rho)} -\frac{1}{2}\right)
\rect\left(\frac{\rho }{2R}\right)
\end{CAlign}

\subsection{Matched Filter}
\label{MFappend}
\label{ReenvelopeMatchedFilterDerivationappend}
With input equal to the motion-free filtered microbubble signal $q(\vc{r}, \vc{\Delta v})$, the matched filter output is
\begin{CAlign}
\label{MFappendpre}
    \psi(\vc{r}, \delv) &\defsign q(\vc{r}, \vc{\Delta v})\ast_{\vc{r}} g( - \vc{r}) \\
  &=   \int\limits_{- \infty}^{\infty}R_g(\vc{r} - t \vc{\Delta v})  w(t ) dt  \label{eq:psi1},
\end{CAlign}
where \eqref{eq:psi1} follows by inserting \eqref{eq:q-g-conv-h} into \eqref{MFappendpre},
where $R_g(\vc{r}) \defsign g(  \vc{r}) \ast_{\vc{r}} g(\vc{r})$ is the autocorrelation function of $g$. Using \eqref{passbandPSF}, it can be shown that
 \begin{align*}
 R_g(\vc{r}) 
 & =\frac{1}{2}\hat{g}_e(\vr)\cos\left(\frac{2\pi}{\lambda} z\right)
 +C_g
  \hat{g}_e(\vr)\\
 \hat{g}_e(\vr) &\defsign  1/(2 \pi  \hat{\sigma}_r^2)
\exp\left(-\|\vc{r} \|^2/(2\hat{\sigma}_r^2) \right) \\
C_g & \defsign\frac{1}{2}\exp\left(-4 \pi^2\frac{\sigma_r^2}{\lambda^2}\right) \\
 \hat{\sigma}_r &\defsign \sqrt{2} \sigma_r
 \end{align*}
Because the two terms in $R_g(\vc{r})$ have the same form as $g(\vr)$ and $g_e(\vr)$ in 
\eqref{passbandPSF} and \eqref{basebandPSF}, respectively, the results of \eqref{eq:q-g-conv-h} are easily translated to those for \eqref{eq:psi1}. In particular,
\begin{CAlign}
    \psi(\vr,\delv) = \psi_1(\vr,\delv)+ \psi_2(\vr,\delv)
\end{CAlign}
where $\psi_1(\vr)$ and $\psi_2(\vr)$ have the same forms as $q(\vr, \delv)$ in \eqref{finalqforI} and in \eqref{motionfreeappendixpost}, respectively,
\begin{CAlign}
\psi_1(\vr)
&= \hat{\Gamma}(\delv) g_e(\hat{\eta} \vc{r}) \cos \left(\frac{ 2\pi }{ \lambda} (z - \hat{\zeta}( \vc{r},\delv)) \right) \\
\psi_2(\vr) & = \frac{2C_g}{\sqrt{4 + 2\kappa^2}}
g_e\left(\hat{\eta}\vc{r}\right) 
\end{CAlign}
where
\vspace*{-4mm}
\begin{CAlign} \label{eq:Gammahat}
    \hat{\Gamma}(\delv) & \defsign 
    \frac{1}{\sqrt{4 + 2\kappa^2}} \exp\left[- \frac{4\pi^2}{2+ \kappa^2} \left(\frac{  \sigma_t {\Delta v}_z}{\lambda }
\right)^2 \right] 
\end{CAlign}
with similar changes from $\eta$ and $\zeta(\vr,\delv)$ to $\hat{\eta}$ and $\hat{\zeta}(\vr,\delv)$.
However, we will only be interested in the peak $\psi(\vc{0}, \delv)$ of the matched filter output at $\vr=0$, at which point $\hat{\zeta}(\vr,\delv)=\hat{\eta} \vr =0$, so $\hat{\eta}$ and $\hat{\zeta}$ become irrelevant, and we omit the expressions for them. It follows that 
\begin{CAlign}
\psi(\vc{0},\delv) = 
\left[\hat{\Gamma}(\delv) + \frac{2C_g}{\sqrt{4 + 2\kappa^2}}\right] g_e(\vc{0})
\end{CAlign}

\subsection{Velocity Filter Bandwidth}
\label{deltavdis}
We now determine the velocity filter bandwidth $\delta v$ defined in Section \ref{secapparentmicrobubblechapter4}. 
The idea described in \eqref{rangeofvelocities}, is to determine $\delta v$ in terms of a classical ``half max''  criterion by thresholding the matched filter output at half of the auto correlation peak, that is, we solve for $\delv$ satisfying
\begin{align*}
\psi(\vc{0},\delv) &= 0.5 \psi(\vc{0},\vc{0}) \\
\hat{\Gamma}(\delv) + \frac{2C_g}{\sqrt{4 + 2\kappa^2}} &= 
0.5\hat{\Gamma}(\vc{0})+ 0.5C_g \\
\hat{\Gamma}(\delv) &= 0.25 + C_g \left[0.5 - (1+  \kappa^2/2)^{-1/2} \right]
\end{align*}
Because $C_g \ll 0.25$ for all reasonable parameter values, the half-max condition simplifies to $\hat{\Gamma}(\delv) =0.25$.

To express the dependence of $\hat{\Gamma}(\delv)$ \eqref{eq:Gammahat} on both magnitude and direction of $\delv$, we parametrize $\delta v = \|\delv\|$, $v_z = \sin(\theta) \delta v$ in the half-max condition, where $\theta$ defines the angle of $\vc{\Delta v}$ with respect to the lateral axis.
Assuming $\sigma_r = \lambda$, the half-max condition can be expressed in terms of  
\begin{CAlign}
\label{definitionkappadelta}
  \kappa_{\delta v} \triangleq \sigma_t \delta v/\sigma_r  
\end{CAlign}
\begin{CAlign}
\label{PsiMFpre2}
 \text{as:} \quad \frac{1}{\sqrt{4 + 2\kappa_{\delta v}^2}}\exp\left(- \frac{4\pi^2\kappa_{\delta v}^2 \sin^2\theta}{2+ \kappa_{\delta v}^2} \right) =0.25
\end{CAlign}
For each fixed $\theta$, \eqref{PsiMFpre2} can be solved numerically for $\kappa_{\delta v}^2$, and hence for
$\delta v$, yielding the values shown in Fig. \ref{deltavfig} in the main text. 
Thanks to \eqref{definitionkappadelta}, $\delta v \propto \sigma_r/\sigma_t$, but the dependence on $\theta$ is nonlinear.
\subsection{Transverse Oscillation -- TO}
\subsubsection{TO PSF}
\label{derivationTOPSF}
The TO PSF is found by taking the inverse Fourier transform $g^{TO}(\vc{r}) = F^{-1}_{\vc{k}\rightarrow \vc{r}}\{G(\vc{k})G_T(\vc{k}) \}$ where $G(\vc{k})$ is the k-space pre-envelope PSF  \eqref{passbandPSF} and 
$G_T(\vc{k})$ is defined in \eqref{TOfilterspec}. Defining $\alpha \defsign \frac{2 \pi}{\lambda}$, $\vc{e}_1\defsign [1,0]^T$, and $\vc{e}_2 \defsign [0,1]^T$,
\begin{align*}
    G(\vc{k})
    &= 0.5 e^{-\frac{\| \vc{k}- \alpha \vc{e}_2 \|^2 \sigma_r^2}{2}} + 0.5e^{-\frac{\| \vc{k}+\alpha \vc{e}_2 \|^2 \sigma_r^2}{2}}  \\
     G_T(\vc{k}) &=e^{-\sigma_{x}^2\left(\vc{k}\cdot \vc{e}_1-k_{0x}\right)^2/2} 
    + e^{-\sigma_{x}^2\left(\vc{k}\cdot \vc{e}_1+k_{0x}\right)^2/2} 
\end{align*}
To simplify, let
\begin{equation*}
M(\vc{k}; \alpha,\mu) \defsign (2 \pi)^2e^{-{0.5\sigma_r^2\| \vc{k}- \alpha \vc{e}_2 \|^2 }} e^{{-0.5\sigma_{x}^2\left(\vc{k}\cdot \vc{e}_1-\mu\right)^2}}
\end{equation*}
with inverse Fourier transform $m(\vc{r};\alpha, \mu)$.
Then 
\begin{CAlign}
\label{eq:GTO-general2}
g^{TO}(\vc{r})= \frac{0.5}{(2 \pi)^2} \sum_{\chi\in \{-1, 1\}} \sum_{\xi \in \{-1, 1\}} m(\vc{r}; \chi\alpha, \xi k_{0x})
\end{CAlign}
Now, 
\begin{equation*}
m(\vc{r};\alpha, \mu) = \int_{\Rtwo} e^{-\| \vc{k}-\alpha \vc{e}_2 \|^2 \sigma_r^2/2}
e^{{-\sigma_{x}^2\left(\vc{k}\cdot \vc{e}_1-\mu\right)^2/2}} e^{j \vc{k}\cdot \vc{r}} d\vc{k}
\end{equation*}
Combining the exponents in the integrand, the resulting exponent is a quadratic form in $\vc{k}$ 
\begin{align*}
a \| \vc{k}-\vc{\beta} \|^2 +\gamma \left(\vc{k}^T \vc{e}_1-\mu\right)^2- & j \vc{k}^T \vc{r} \\
& = (\vc{k}-\vc{b})^T A (\vc{k}-\vc{b}) + f \nonumber
\end{align*}     
where $a \defsign \sigma_r^2/2$, $\vc{\beta} \defsign \alpha \vc{e}_2$, $\gamma \defsign \sigma_x^2/2$, for some constant matrix $A$, vector $\vc{b}$ and scalar $f$, that are to be determined.

It follows that
\begin{CAlign}
m(\vc{r};\alpha, \mu)
&= \exp(-f) \int_{\Rtwo} \exp\left[-(\vc{k}-\vc{b})^T A(\vc{k}-\vc{b})\right] d\vc{k} \nonumber \\
&=\pi  \exp(-f) \det(A)^{-1/2} \label{eq:m-one}
\end{CAlign}
where \eqref{eq:m-one} follows by
recognizing the integrand as an unormalized
 2D Gaussian density.
To proceed, we determine the coefficients in the quadratic form as follows.
\[A  = aI + \gamma \vc{e}_1 \vc{e}_1^T
\]
with determinant and inverse given by the matrix determinant lemma and the Sherman-Morrison Formula, respectively as
\begin{align*}
det(A) &= a^2 [1 +(\gamma/a)\vc{e}_1^T \vc{e}_1]  
= a^2 + \gamma a\\
A^{-1} 
&= 
(1/a)I -\gamma/(a^2+\gamma a )\vc{e}_1 \vc{e}_1^T
\end{align*}
and,
\begin{align*}
\vc{b} &=  A^{-1}(a \vc{\beta}+ \gamma \mu \vc{e}_1 + (j/2) \vc{r}) \\
f &= \gamma \mu^2 + a \alpha^2 -\vc{b}^T A \vc{b} \\
&=  \mu^2 a\gamma/(\gamma+a) + 0.25/(\gamma+a)x^2+0.25/az^2 \\
& \quad - j \mu \gamma/  (a+\gamma) x -j \alpha z
\end{align*}
Substituting into \eqref{eq:m-one} yields
\begin{CAlign}
\label{eq:m-final2}
m(\vc{r};\alpha, \mu) = 
C_1(\mu)
\exp\left[
-\vc{r}^T D \vc{r}/2 +j \vc{\zeta}(\alpha,\mu)^T \vc{r}\right]
\end{CAlign}
where
\begin{align*}
    C_1(\mu) & \defsign \frac{\pi}{a \sqrt{(1 +(\gamma/a)}}
\exp\left(
-\frac{a  \gamma \mu^2}{\gamma+a} 
     \right) \\
     &= \frac{2 \pi}{\sigma_r^2 \sqrt{(1 +(\sigma_x^2/\sigma_r^2)}}
\exp\left[
-\frac{ 2 \pi^2 \sigma_r^2}{\tilde{\lambda}_x^2} \left(1+ \frac{\sigma_r^2}{\sigma_x^2}\right)
\right] \\
     D & \defsign 0.5\diag\left[1/(\gamma+a),  1/a \right] \\
     & = \diag\left[1/(\sigma_r^2+\sigma_x^2),  1/\sigma_r^2 \right] \\
     \zeta(\mu,\alpha)^T & \defsign (\mu \gamma/  (a+\gamma)  ,\alpha) = \begin{bmatrix} \mu \sigma_r^2/  (\sigma_r^2 + \sigma_x^2)  &\alpha \end{bmatrix}
\end{align*}
Combining \eqref{eq:GTO-general2} and \eqref{eq:m-final2} yields the final expression 
\begin{align*}
 g^{TO}(\vc{r})
 &= 0.5/(2 \pi)^2 C_1(\mu) \exp(-\vc{r}^T D \vc{r}/2) \nonumber \\
 &\times [ 2\cos\left(  \vc{\zeta}(\alpha,\mu)^T \vc{r}\right)+ 2\cos\left( j \vc{\zeta}(-\alpha,\mu)^T \vc{r}\right) ] \nonumber
 \\
 &= 
 g^{TO}_e(\vc{r}) \cos\left(\frac{\gamma k_{0x}}{(a+\gamma)}  x \right) \cos\left(\alpha z\right)
\end{align*}
where the envelope is
\begin{align*}
g^{TO}_e(\vc{r})
&=(4 \pi \sigma_r^2) /(2\pi)^2 C_1(k_{0x}) 
g_e\left( \sigma_r D^{1/2} \vc{r} \right)
\end{align*}
Finally, substituting the definitions of the various constants, we obtain
\begin{CAlign} \label{eq:gTO-final}
g^{TO}(\vc{r}) =  g^{TO}_e(\vc{r}) \cos\left(\frac{2\pi x}{\tilde{\lambda}_x} \right) \cos\left(\frac{2\pi z}{\lambda} \right)
\end{CAlign}
where
\begin{CAlign}
\tilde{\lambda}_x &\defsign (1+\sigma_r^2/\sigma_x^2) \lambda_x \nonumber \\
\label{eq:gTO-env-final}
g^{TO}_e(\vc{r}) &=C^{TO}g_e\left(\left[ (1+\sigma_x^2/\sigma_r^2)^{-1/2}, 1 \right] \vc{r} \right) \\
C^{TO} &\defsign \frac{2}{ \sqrt{(1 +\sigma_x^2/\sigma_r^2)}}
\exp\left[
-\frac{ 2 \pi^2 \sigma_r^2}{\tilde{\lambda}_x^2} \left(1+ \frac{\sigma_r^2}{\sigma_x^2}\right)
\right] \nonumber
\end{CAlign}

\subsubsection{Velocity Filtering with TO}
\label{appendTOimp}
To obtain $q^{TO}(\vc{r};\vc{\Delta v})$ we use \eqref{eq:q-g-conv-h}, replacing $g(\vc{r})$ by $g^{TO}(\vc{r})$ from \eqref{eq:GTO-general2}.
%

Let 
\begin{CAlign}
 p(\vc{r},\vc{\Delta v};\alpha, \mu)  & \defsign
 \int_{- \infty}^{\infty} m(\vc{r}-t\vc{\Delta v};\alpha, \mu)   w(t ) dt  \label{eq:p-def} \\ 
 & = \frac{m(\vc{r};\alpha, \mu)}{\sqrt{2\pi \sigma_t^2}}       \int\limits_{- \infty}^{\infty} \exp(-a_2t^2/2+a_1t) dt \nonumber \\
 & = \frac{m(\vc{r};\alpha, \mu)}{\sqrt{\sigma_t^2 a_2}} \exp\left(\frac{a_1^2}{2a_2}\right) \label{eq:p}
\end{CAlign}
      where the last equality is obtained using a standard integral expression and where 
      \begin{align*}
          a_1 & \defsign \vc{\Delta v} ^T D \vc{r} -j \vc{\zeta(\mu,\alpha)}^T \vc{\Delta v} \\ 
          a_2 & \defsign  \vc{\Delta v} ^T D \vc{\Delta v} + 1/\sigma_t^2 
      \end{align*}
  Substituting $a_1$ and $a_2$ into \eqref{eq:p} yields
  \begin{CAlign}
      p(\vc{r},\vc{\Delta v}; \mu) &=
      \Gamma(\vc{\Delta v}; \mu, \alpha)
    \Lambda(\vc{r},\vc{\Delta v};\mu)
e^{j \zeta(\mu,\alpha)^T
J \vc{r}}
\label{eq:p-two}
  \end{CAlign}
  where
\begin{align*}
  \Gamma(\vc{\Delta v}; \mu, \alpha)
  &\defsign \frac{1}{\sqrt{\tilde{\kappa}^2 + 1}}
\exp\left(-\frac{\sigma_t^2 [\vc{\zeta(\mu,\alpha)}^T \vc{\Delta v}]^2
}{2(\tilde{\kappa}^2 +1)} \right) \nonumber\\
   \Lambda(\vc{r},\vc{\Delta v}; \mu)
      & \defsign C_1(\mu)
\exp\left(-\frac{\vc{r}^T D \vc{r}}{2} 
+\frac{\sigma_t^2(\vc{\Delta v} ^T D \vc{r})^2}
{2(\tilde{\kappa}^2 +1)}\right) \\
& =2 \pi^2g_e^{TO}\left[ \Xi(\delv) \vc{r} \right]  \nonumber \\
\Xi(\vc{\Delta v}) & \defsign I - \frac{ \sigma_t^2}{\tilde{\kappa}^2} \left(1- \frac{1}{\sqrt{1+\tilde{\kappa}^2}}
    \right)\vc{\Delta v} \vc{\Delta v}^T D \nonumber\\
J &\defsign
I - \sigma_t^2/(\tilde{\kappa}^2 +1) \vc{\Delta v} \vc{\Delta v}^T D  \nonumber\\
  \tilde{\kappa}^2 & \defsign \sigma_t^2\vc{\Delta v} ^T D \vc{\Delta v}
  \nonumber 
  \end{align*}
  Next, using \eqref{eq:q-g-conv-h}, \eqref{eq:GTO-general2}, 
  \eqref{eq:p-def}, and \eqref{eq:p-two} we have
   \begin{CAlign}
   \label{eq:q-sum}
       q(\vc{r};\vc{\Delta v}) & = 
 0.5/(2 \pi)^2 \sum_{\chi\in \{-1, 1\}} \sum_{\xi \in \{-1, 1\}} p(\vc{r}; \chi\alpha, \xi k_{0x}) \nonumber \\
 & = g_e^{TO}[\Xi(\vc{\Delta v}) \vc{r} ] \left\{ \Gamma_1 \cos\theta_1 +\Gamma_2 \cos \theta_2 \right\}
\end{CAlign}
where, for $j=1,2 $,
\begin{align*}
\Gamma_j & \defsign 
\Gamma(\vc{\Delta v}; (-1)^j k_{0x}, \alpha)/2
\\
&= \frac{1}{2 \sqrt{1 + \tilde{\kappa}^2 }} \exp \left[- \frac{2\pi^2 \sigma_t^2}{1+ \tilde{\kappa}^2} \left(\frac{   {\Delta v}_z}{\lambda_z } - \frac{ (-1)^j    \Delta v_x}{\tilde{\lambda_x} } \right)^2 \right] 
\\
\theta_j & \defsign  \zeta[(-1)^j k_{0x},\alpha]^T J \vc{r}\\
 &=\frac{2\pi z}{\lambda_z} 
 - (-1)^j \frac{2\pi x}{\tilde{\lambda_x}}  -\left(\frac{2\pi \sigma_{t}^2}{1 + \tilde{\kappa}^2}\right)  
\left( \frac{{\Delta v}_z}{\lambda_z} - \frac{(-1)^j {\Delta v}_x}{\tilde{\lambda_x}} \right) \\
& \quad \times
\vc{\Delta v}^T D \vc{r}
\end{align*}

\subsection{Noise Analysis}
\label{appendNoise}
To avoid signal loss, the standard processing chain without velocity filtering would need to use the full spatio-temporal passband i.e., the support of $B(\vk, \Omega)$;  we assume it bandlimits the signal to this support, to limit the noise power.
With the flat noise spectrum model described in Sec.~\ref{noiseanalysis}, the power spectral density (PSD) $S_n(\vk,\Omega)$ of the noise  is then
\[ S_n (\vk,\Omega) = \begin{cases} 
     N_0 & \|\vk\| \leq k_G |, |\Omega| \leq \Omega_{\mathrm{max}} = k_G v_{0,\max} \\
      0 & \mathrm{otherwise}
   \end{cases}
\]
The average power $\mathbf{E}[(n)^2(\vr,t)] $ of the input noise to the localization step without velocity filtering is:
\begin{CAlign}
   \mathbf{E}[(n)^2(\vr,t)] &= \frac{1}{(2\pi)^3}\int_{\mathbb{R}^3}  S_n (\vk,\Omega)   d\vk d\Omega \nonumber \\ 
   &= N_0 \frac{\Omega_{\mathrm{max}}
   k_G^2}{4 \pi^{2}} 
 \end{CAlign} In contrast, after the velocity filter, the   noise power is:
\begin{CAlign}
\qquad & \hspace*{-10mm} \mathbf{E}[(n')^2(\vr,t)] \nonumber \\
    &= \frac{1}{(2\pi)^3}\int_{\mathbb{R}^3} S_n (\vk,\Omega) |H( \vk,\Omega)|^2   d\vk d\Omega \nonumber \\ 
   &= \frac{ N_0}{(2\pi)^3} \int_{ || \vk|| < k_G } d \vk \int_{-\Omega_{\mathrm{max}}}
   ^{\Omega_{\mathrm{max}}}
   e^{- \sigma_t^2(\Omega + \vc{k} \cdot \vc{v_0})^2} d\Omega \nonumber
   \\
   & \leq \frac{N_0}{(2\pi)^3} \int_{ || \vk|| < k_G } d \vk \int_{-\infty}
   ^{\infty}
    e^{- \sigma_t^2(\Omega + \vc{k} \cdot \vc{v_0})^2} d\Omega \nonumber \\
   &= N_0 \frac{k_G^2}{8 \pi^{1.5} \sigma_t}  
\end{CAlign}
It follows that noise power is reduced by the velocity filter by the factor
\begin{CAlign}
NRF & \defsign\frac{\mathbf{E}[(n)^2(\vr,t)]}{\mathbf{E}[(n')^2(\vr,t)]} \nonumber \\
& \geq
\frac{2}{\sqrt{\pi}} \Omega_{\mathrm{max}} \sigma_t = \frac{2}{\sqrt{\pi}} k_G v_{0,\max} \sigma_t
\end{CAlign}

While the analysis here is performed in the continuous $(\vk,\Omega)$ domain, the system operates on individual snapshots of the data, sampled at the frame rate of $F=1/T$. If the frame rate is chosen to the Nyquist rate to avoid aliasing of the spatio-temporal data of the moving bubbles, then $\Omega_{\mathrm{max}} = \pi F$, and the NRF may be expressed in terms of $F$ as
\[
NRF \geq 2 \sqrt{\pi} \sigma_t F
\]

\end{document}